\documentclass[ALICE,manyauthors]{cernphprep}
\usepackage[comma,square,numbers,sort&compress]{natbib}
\usepackage{hyperref}
\usepackage{lineno}
\usepackage{xspace}
\usepackage{array,booktabs}
\usepackage{multirow}
\usepackage[T1]{fontenc} 

\begin{document}
\newcommand {\red}[1]    {\textcolor{red}{#1}}
\newcommand {\green}[1]  {\textcolor{green}{#1}}
\newcommand {\blue}[1]   {\textcolor{blue}{#1}}
\newcommand {\dgreen}[1]    {\textcolor{darkgreen}{#1}}
\newcommand {\orange}[1]   {\textcolor{orange}{#1}}

\newcommand{\deutsch}[1]{\foreignlanguage{german}{#1}}

\newcommand{\CZ}[1]{\blue{{\textbf{CZ:} #1}}}
\newcommand{\PA}[1]{\red{{\textbf{PA:} #1}}}
\newcommand{\JN}[1]{\red{{\textbf{JN:} #1}}}
\newcommand{\MF}[1]{\red{{\textbf{JN:} #1}}}
\newcommand{\ToDo}      {\textcolor{blue}{\footnotesize \textsc{ToDo}}}
\newcommand{\TBC}       {\textcolor{red}{\footnotesize \textsc{TBC}}}

\DeclareRobustCommand{\unit}[2][]{%
        \begingroup%
                \def\0{#1}%
                \expandafter%
        \endgroup%
        \ifx\0\@empty%
                \ensuremath{\mathrm{#2}}%
        \else%
                \ensuremath{#1\,\mathrm{#2}}%
        \fi%
        }
\DeclareRobustCommand{\unitfrac}[3][]{%
        \begingroup%
                \def\0{#1}%
                \expandafter%
        \endgroup%
        \ifx\0\@empty%
                \raisebox{0.98ex}{\ensuremath{\mathrm{\scriptstyle#2}}}%
                \nobreak\hspace{-0.15em}\ensuremath{/}\nobreak\hspace{-0.12em}%
                \raisebox{-0.58ex}{\ensuremath{\mathrm{\scriptstyle#3}}}%
        \else
                \ensuremath{#1}\,%
                \raisebox{0.98ex}{\ensuremath{\mathrm{\scriptstyle#2}}}%
                \nobreak\hspace{-0.15em}\ensuremath{/}\nobreak\hspace{-0.12em}%
                \raisebox{-0.58ex}{\ensuremath{\mathrm{\scriptstyle#3}}}%
        \fi%
}

%
%
\newcommand{\ie}{i.\,e.\;}
\newcommand{\eg}{e.\,g.\;}

\newcommand{\run}[1]{Run #1}

%
%

\newcommand{\dNdeta}{\ensuremath{\mathrm{d}N_{\rm ch}/\mathrm{d}\eta}\xspace}

%
%
\newlength{\smallerpicsize}
\setlength{\smallerpicsize}{70mm}
\newlength{\smallpicsize}
\setlength{\smallpicsize}{90mm}
\newlength{\mediumpicsize}
\setlength{\mediumpicsize}{120mm}
\newlength{\largepicsize}
\setlength{\largepicsize}{150mm}

\newcommand{\PICX}[5]{
   \begin{figure}[!hbt]
      \begin{center}
         \vspace{3ex}
         \includegraphics[width=#3]{#1}
         \caption[#4]{\label{#2} #5}        
      \end{center}  
   \end{figure}
}

\newcommand{\PICH}[5]{
   \begin{figure}[H]
      \begin{center}
         \vspace{3ex}
         \includegraphics[width=#3]{#1}
         \caption[#4]{\label{#2} #5}        
      \end{center}  
   \end{figure}
}

%
%
%
\newcommand{\figs}{Figs.\xspace}
\newcommand{\Figs}{Figures\xspace}
\newcommand{\eqn}{equation\xspace}
\newcommand{\Eqn}{Equation\xspace}
\newcommand{\figref}[1]{Fig.~\ref{#1}}
\newcommand{\Figref}[1]{Figure~\ref{#1}}
\newcommand{\figrefs}[2]{Figs.~\ref{#1} and~\ref{#2}}
\newcommand{\Figrefs}[2]{Figures~\ref{#1} and~\ref{#2}}
\newcommand{\tabref}[1]{Tab.~\ref{#1}}
\newcommand{\Tabref}[1]{Table~\ref{#1}}
\newcommand{\appref}[1]{appendix~\ref{#1}}
\newcommand{\Appref}[1]{Appendix~\ref{#1}}
\newcommand{\secs}{Secs.\xspace}
\newcommand{\Secs}{Sections\xspace}
\newcommand{\secref}[1]{Sec.~\ref{#1}}
\newcommand{\Secref}[1]{Section~\ref{#1}}
\newcommand{\chaps}{Chaps.\xspace}
\newcommand{\Chaps}{Chapters\xspace}
\newcommand{\chapref}[1]{Chap.~\ref{#1}}
\newcommand{\Chapref}[1]{Chapter~\ref{#1}}
\newcommand{\lstref}[1]{Listing~\ref{#1}}
\newcommand{\Lstref}[1]{Listing~\ref{#1}}
%
%
\newcommand{\otoprule}{\midrule[\heavyrulewidth]}
\topfigrule
%
\newcommand {\stat}     {({\it stat.})~}
\newcommand {\syst}     {({\it syst.})~}
 \newcommand {\mom}       {\ensuremath{p}}
\newcommand {\pT}        {\pt}
\newcommand {\meanpT}    {\ensuremath{\langle p_{\mathrm{T}} \kern-0.1em\rangle}\xspace}
\newcommand {\mean}[1]   {\ensuremath{\langle #1 \kern-0.1em\rangle}\xspace} 
\newcommand {\sqrtsNN}   {\ensuremath{\sqrt{s_{\textsc{NN}}}}\xspace}
\newcommand {\sqrts}     {\ensuremath{\sqrt{s}}\xspace}
\newcommand {\vf}        {\ensuremath{v_{\mathrm{2}}}\xspace}
\newcommand {\et}        {\ensuremath{E_{\mathrm{t}}}\xspace}
\newcommand {\mT}        {\ensuremath{m_{\mathrm{T}}}\xspace}
\newcommand {\mTmZero}   {\ensuremath{m_{\mathrm{T}} - m_0}\xspace}
\newcommand {\minv}      {\mbox{$m_{\ee}$}}
\newcommand {\rap}       {\mbox{$y$}}
\newcommand {\abs}[1]    {\ensuremath{\left | #1 \right |}}
\newcommand {\absrap}    {\mbox{$\left | y \right | $}}
\newcommand {\rapXi}     {\mbox{$\left | y(\rmXi) \right | $}}
\newcommand {\abspseudorap} {\mbox{$\left | \eta \right | $}}
\newcommand {\pseudorap} {\mbox{$\eta$}}
\newcommand {\cTau}      {\ensuremath{c\tau}}
\newcommand {\sigee}     {$\sigma_E$/$E$}
\newcommand {\dNdy}      {\ensuremath{\mathrm{d}N/\mathrm{d}y}}
\newcommand {\dNdpt}     {\ensuremath{\mathrm{d}N/\mathrm{d}\pT }}
\newcommand {\dNdptdy}   {\ensuremath{\mathrm{d^{2}}N/\mathrm{d}\pT\mathrm{d}y }}
\newcommand {\fracdNdptdy}   {\ensuremath{ \frac{\mathrm{d^{2}}N}{\mathrm{d}\pT\mathrm{d}y } }}
\newcommand {\dNdmtdy}   {\ensuremath{\mathrm{d^{2}}N/\mathrm{d}\mT\mathrm{d}y }}
\newcommand {\dN}        {\ensuremath{\mathrm{d}N }}
\newcommand {\dNsquared} {\ensuremath{\mathrm{d^{2}}N }}
\newcommand {\dpt}       {\ensuremath{\mathrm{d}\pT }}
\newcommand {\dy}        {\ensuremath{\mathrm{d}y}}
\newcommand {\dNdyBold}  {\ensuremath{\boldsymbol{\dN/\dy}}\xspace}
\newcommand {\dNchdy}    {\ensuremath{\mathrm{d}N_\mathrm{ch}/\mathrm{d}y }\xspace}
\newcommand {\dNchdeta}  {\ensuremath{\mathrm{d}N_\mathrm{ch}/\mathrm{d}\eta }\xspace}
\newcommand {\dNchdptdeta}  {\ensuremath{\mathrm{d}N_\mathrm{ch}/\mathrm{d}\pT\mathrm{d}\eta }\xspace}
\newcommand {\Raa}       {\ensuremath{R_\mathrm{AA}}}
\newcommand {\RpPb}       {\ensuremath{R_\mathrm{pPb}}\xspace}
\newcommand {\Nevt}      {\ensuremath{N_\mathrm{evt}}}
\newcommand {\NevtINEL}  {\ensuremath{N_\mathrm{evt}(\textsc{inel})}}
\newcommand {\NevtNSD}   {\ensuremath{N_\mathrm{evt}(\textsc{nsd})}}
\newcommand{\dEdx}       {\ensuremath{\mathrm{d}E/\mathrm{d}x}\xspace}
\newcommand{\ttof}       {\ensuremath{t_\mathrm{TOF}}\xspace}
\newcommand {\ee}        {\mbox{$\mathrm {e^+e^-}$}\xspace}
\newcommand {\ep}        {\mbox{$\mathrm {e^-p}$}\xspace}
\newcommand {\pp}        {\mbox{$\mathrm {pp}$}\xspace}
\newcommand {\ppBoldMath} {\mbox{$\mathrm { \mathbf p \mathbf p }$}\xspace}
\newcommand {\ppbar}     {\mbox{$\mathrm {p\overline{p}}$}\xspace}
\newcommand {\PbPb}      {\ensuremath{\mbox{Pb--Pb}}\xspace}
\newcommand {\AuAu}      {\ensuremath{\mbox{Au--Au}}\xspace}
\newcommand {\CuCu}      {\ensuremath{\mbox{Cu--Cu}}\xspace}
\renewcommand {\AA}      {\ensuremath{\mbox{A--A}}\xspace}
\newcommand {\pA}        {\ensuremath{\mbox{p--A}}\xspace}
\newcommand {\pPb}       {\ensuremath{\mbox{p--Pb}}\xspace}
\newcommand {\Pbp}       {\ensuremath{\mbox{Pb--p}}\xspace}
\newcommand {\hPM}       {\ensuremath{h^{\pm}}\xspace}
\newcommand {\rphi}      {\ensuremath{(r,\phi)}\xspace}
\newcommand {\alphaS}    {\ensuremath{ \alpha_s}\xspace}
\newcommand {\MeanNpart} {\mbox{\ensuremath{< \kern-0.15em N_{part} \kern-0.15em >}}}

\newcommand {\sig}       {\ensuremath{S}\xspace}
\newcommand {\expsig}    {\ensuremath{\hat{S}}\xspace}
\newcommand {\prob}      {\ensuremath{P}\xspace}
\newcommand {\prior}     {\ensuremath{C}\xspace}
\newcommand {\prop}      {\ensuremath{F}\xspace}
\newcommand {\atrue}     {\ensuremath{\vec{A}_{\mathrm{true}}}\xspace}
\newcommand {\ameas}     {\ensuremath{\vec{A}_{\mathrm{meas}}}\xspace}
\newcommand {\detresp}   {\ensuremath{R}\xspace}

\newcommand {\pid}       {\ensuremath{\mathrm{\epsilon}_\mathrm{PID}}\xspace}
\newcommand {\nsigma}    {\ensuremath{\mathrm{n_{\sigma}}}\xspace}
\newcommand {\ylab} {\ensuremath{\mathrm{| y_{lab} |}}\xspace}

%
%
\newcommand {\mass}     {\mbox{\rm MeV$\kern-0.15em /\kern-0.12em c^2$}}
\newcommand {\tev}      {\mbox{${\rm TeV}$}\xspace}
\newcommand {\gev}      {\mbox{${\rm GeV}$}\xspace}
\newcommand {\mev}      {\mbox{${\rm MeV}$}\xspace}
\newcommand {\kev}      {\mbox{${\rm keV}$}\xspace}
\newcommand {\tevBoldMath}  {\mbox{${\rm \mathbf{TeV}}$}}
\newcommand {\gevBoldMath}  {\mbox{${\rm \mathbf{GeV}}$}}
\newcommand {\mmom}     {\mbox{\rm MeV$\kern-0.15em /\kern-0.12em c$}}
\newcommand {\gmom}     {\mbox{\rm GeV$\kern-0.15em /\kern-0.12em c$}}
\newcommand {\mmass}    {\mbox{\rm MeV$\kern-0.15em /\kern-0.12em c^2$}}
\newcommand {\gmass}    {\mbox{\rm GeV$\kern-0.15em /\kern-0.12em c^2$}}
\newcommand {\nb}       {\mbox{\rm nb}}
\newcommand {\musec}    {\mbox{$\mu {\rm s}$}}
\newcommand {\nsec}     {\mbox{${\rm ns}$}}
\newcommand {\psec}     {\mbox{${\rm ps}$}}
\newcommand {\fmC}      {\mbox{${\rm fm/c}$}}
\newcommand {\fm}       {\mbox{${\rm fm}$}}
\newcommand {\cm}       {\mbox{${\rm cm}$}}
\newcommand {\mm}       {\mbox{${\rm mm}$}}
\newcommand {\mim}      {\mbox{$ \mu {\rm m}$}}
\newcommand {\cmq}      {\mbox{${\rm cm}^{2}$}}
\newcommand {\mmq}      {\mbox{${\rm mm}^{2}$}}
\newcommand {\dens}     {\mbox{${\rm g}/{\rm cm}^{3}$}}
\newcommand {\lum}      {\, \mbox{${\rm cm}^{-2} {\rm s}^{-1}$}}
\newcommand {\barn}     {\, \mbox{${\rm barn}$}}
\newcommand {\m}        {\, \mbox{${\rm m}$}}
\newcommand {\dg}       {\mbox{$\kern+0.1em ^\circ$}}
\newcommand{\mpp}{\ensuremath{\mathrm{pp}}\xspace}
\newcommand{\rts}{\ensuremath{\sqrt{s}}\xspace}
\newcommand{\GeV}{\ensuremath{\mathrm{GeV}}\xspace}
\newcommand{\TeV}{\ensuremath{\mathrm{TeV}}\xspace}
\newcommand{\gevc}{\ensuremath{\mathrm{GeV}/c}\xspace}
\newcommand{\GeVc}{\gevc}
\newcommand{\mevc}{\ensuremath{\mathrm{MeV}/c}\xspace}
\newcommand{\mevcc}{\ensuremath{\mathrm{MeV}/c^{2}}\xspace}
\newcommand{\gevcc}{\ensuremath{\mathrm{GeV}/c^{2}}\xspace}
\newcommand{\pt}{\ensuremath{p_{\rm T}}\xspace}
\newcommand{\kt}{\ensuremath{k_{\rm T}}\xspace}
\newcommand {\lumi}{\mathcal{L}_{\rm int}\xspace}
\newcommand{\nbinv}{\ensuremath{\rm nb^{-1}}}
\newcommand {\ubinv}{\ensuremath{\mu\rm b^{-1}}}
\newcommand {\um}{\ensuremath{\mu\rm m}\xspace}
\newcommand{\mub}{\ensuremath{\mu\rm b}\xspace}
\newcommand{\lt}{\textless}
\newcommand{\ctau}{\ensuremath{c\tau}\xspace}

%
%

\newcommand{\ePlusMinus}       {\mbox{$\mathrm {e^{\pm}}$}\xspace}
\newcommand{\muPlusMinus}      {\mbox{$\mathrm {\mu^{\pm}}$}\xspace}

\newcommand{\pion}            {\mbox{$\mathrm {\pi}$}\xspace}
\newcommand{\piZero}            {\mbox{$\mathrm {\pi^0}$}\xspace}
\newcommand{\piMinus}           {\ensuremath{\mathrm {\pi^-}}\xspace}
\newcommand{\piPlus}            {\ensuremath{\mathrm {\pi^+}}\xspace}
\newcommand{\piPlusMinus}       {\mbox{$\mathrm {\pi^{\pm}}$}\xspace}

\newcommand{\proton}    {\mbox{$\mathrm {p}$}\xspace}
\newcommand{\pbar}      {\mbox{$\mathrm {\overline{p}}$}\xspace}
\newcommand{\pOuPbar}   {\mbox{$\mathrm {p^{\pm}}$}\xspace}
\newcommand{\DZero}     {\ensuremath{\mathrm {D^0}}\xspace}
\newcommand{\DZerobar}  {\ensuremath{\mathrm {\overline{D}^0}}\xspace}
\newcommand{\Bminus}    {\ensuremath{\mathrm {B^-}}\xspace}
\newcommand{\BZero}     {\ensuremath{\mathrm {B^0}}\xspace}
\newcommand{\BZerobar}  {\ensuremath{\mathrm {\overline{B}{}^0}}\xspace}
\newcommand{\Bs}     {\ensuremath{\mathrm {B^0_s}}\xspace}
\newcommand{\Bsbar}  {\ensuremath{\mathrm {\overline{B}{}^0_s}}\xspace}

\newcommand{\Dmes}       {\ensuremath{\mathrm {D}}\xspace}

\newcommand{\Lb}{\ensuremath{\rm {\Lambda_b^{0}}}\xspace}
\newcommand{\Xic}         {\ensuremath{\mathrm {\Xi_{c}}}\xspace}
\newcommand{\lambdab}     {\ensuremath{\mathrm {\Lambda_{b}^{0}}}\xspace}
\newcommand{\lambdac}     {\ensuremath{\mathrm {\Lambda_{c}^{+}}}\xspace}
\newcommand{\xicz}        {\ensuremath{\mathrm {\Xi_{c}^{0}}}\xspace}
\newcommand{\xiczp}        {\ensuremath{\mathrm {\Xi_{c}^{0,+}}}\xspace}
\newcommand{\xicp}        {\ensuremath{\mathrm {\Xi_{c}^{+}}}\xspace}
\newcommand{\xib}        {\ensuremath{\mathrm {\Xi_{b}}}\xspace}
\newcommand{\LambdaParticle}        {\ensuremath{\mathrm {\Lambda}}\xspace}

\newcommand{\rmLambdaZ}         {\ensuremath{\mathrm {\Lambda}}\xspace}
\newcommand{\rmAlambdaZ}        {\ensuremath{\mathrm {\overline{\Lambda}}}\xspace}
\newcommand{\rmLambda}          {\ensuremath{\mathrm {\Lambda}}\xspace}
\newcommand{\rmAlambda}         {\ensuremath{\mathrm {\overline{\Lambda}}}\xspace}
\newcommand{\rmLambdas}         {\ensuremath{\mathrm {\Lambda \kern-0.2em + \kern-0.2em \overline{\Lambda}}}\xspace}

\newcommand{\Vzero}             {\ensuremath{\mathrm {V^0}}\xspace}
\newcommand{\Vzerob}             {\ensuremath{{\bold \mathrm {V^0}}}\xspace}
\newcommand{\Kzero}             {\ensuremath{\mathrm {K^0}}\xspace}
\newcommand{\Kzs}               {\ensuremath{\mathrm {K^0_S}}\xspace}
\newcommand{\phimes}            {\ensuremath{\mathrm {\phi}}\xspace}
\newcommand{\Kminus}            {\ensuremath{\mathrm {K^-}}\xspace}
\newcommand{\Kplus}             {\ensuremath{\mathrm {K^+}}\xspace}
\newcommand{\Kstar}             {\ensuremath{\mathrm {K^*}}\xspace}
\newcommand{\Kplusmin}          {\ensuremath{\mathrm {K^{\pm}}}\xspace}
\newcommand{\Jpsi}              {\ensuremath{\rm J/\psi}\xspace}
\newcommand{\DtoKpi}{\ensuremath{\rm D^0\to K^-\pi^+}\xspace}
\newcommand{\DtoKpipi}{\ensuremath{\rm D^+\to K^-\pi^+\pi^+}\xspace}
\newcommand{\DstartoDpi}{\ensuremath{\rm D^{*+}\to D^0\pi^+}\xspace}
\newcommand{\Dzero}{\ensuremath{\mathrm {D^0}}\xspace}
\newcommand{\Dzerobar}{\ensuremath{\mathrm{\overline{D}^0}}\xspace}
\newcommand{\Dstar}{\ensuremath{\rm D^{*+}}\xspace}
\newcommand{\Dplus}{\ensuremath{\rm D^+}\xspace}
\newcommand{\Ds}{\ensuremath{\rm D_s^+}\xspace}
\newcommand{\Dsubs}{\ensuremath{\rm D_{s}^+}\xspace}
\newcommand{\decleng}{\ensuremath{\rm L}_{xyz}}
\newcommand{\Lcminus}{\ensuremath{\rm {\overline{\Lambda}{}_c^-}}\xspace}
\newcommand{\Lcplus}{\lambdac}
\newcommand{\Lc}         {\Lcplus}
\newcommand{\LcD} {\ensuremath{\lambdac/\Dzero}\xspace}

\newcommand{\Lbzero}{\ensuremath{\rm {\Lambda_b^0}}\xspace}
\newcommand{\LctopKpi}{\ensuremath{\rm \Lambda_{c}^{+}\to p K^-\pi^+}\xspace}
\newcommand{\LcpKpi}{\LctopKpi}
\newcommand{\LbtoLc}{\ensuremath{\rm \Lambda_{b}^{0}\to \Lc + \rm{X}}\xspace}
\newcommand{\LctopKzS}{\ensuremath{\rm \Lambda_{c}^{+}\to p K^{0}_{S}}\xspace}
\newcommand{\LcpKs}{\LctopKzS}
\newcommand{\LctoenuLambda}{\ensuremath{\rm \Lambda_{c}^{+}\to e^{+} \nu_{e} \Lambda}\xspace}
\newcommand{\cosP}{\ensuremath{\rm cos_{\Theta_{pointing}}}\xspace}
\newcommand{\KzStopippim}{\ensuremath{\rm K^{0}_{S}\to \pi^{+} \pi^{-}}\xspace}
\newcommand{\Lambdatoppim}{\ensuremath{\rm \Lambda \to p \pi^{-}}\xspace}
\newcommand{\nue}{$\nu_e$}
\newcommand{\DtopiKzs}{\ensuremath{\rm D^+\to \pi^+ K^{0}_{S}}\xspace}
\newcommand{\DstoKKzs}{\ensuremath{\rm D_s^+\to K^+ K^{0}_{S}}\xspace}

\newcommand{\ptLc}{\ensuremath{p_{\rm T, \Lambda_c}}\xspace}
\newcommand{\ptpion}{\ensuremath{p_{\rm T, \pi}}\xspace}
\newcommand{\ptK}{\ensuremath{p_{\rm T, K}}\xspace}
\newcommand{\ptproton}{\ensuremath{p_{\rm T, \proton}}\xspace}

\newcommand{\sqrtsseven}{\ensuremath{\sqrt{s} = 7~\TeV}\xspace}
\newcommand{\sqrtsfive}{\ensuremath{\sqrt{s} = 5.02~\TeV}\xspace}
\newcommand{\sqrtsNNfive}{\ensuremath{\sqrt{s_\mathrm{NN}} = 5.02~\TeV}\xspace}
\newcommand{\sqrtsNNtwohun}{\ensuremath{\sqrt{s_\mathrm{NN}} = 200~\GeV}\xspace}

\newcommand{\Pythia}{{\sc pythia}}
\newcommand{\Powheg}{{\sc powheg}}
\newcommand{\Powlang}{{\sc powlang}}
\newcommand{\Dipsy}{{\sc dipsy}}
\newcommand{\Herwig}{{\sc Herwig}}
\newcommand{\Fonl}{{\sc fonll}}

\begin{titlepage}
\PHyear{2020}       
\PHnumber{218}      
\PHdate{10 November}  

\title{\Lcplus~production in pp and in \pPb collisions at \sqrtsNN = 5.02~TeV}
\ShortTitle{\Lcplus~production in ALICE}   

\Collaboration{ALICE Collaboration\thanks{See Appendix~\ref{app:collab} for the list of collaboration members}}
\ShortAuthor{ALICE Collaboration} 

\begin{abstract}

The production cross section of prompt \Lcplus charm baryons was measured with the ALICE detector at the 
LHC at midrapidity in proton-proton (\pp) and proton-lead (\pPb) collisions at a centre-of-mass energy per nucleon pair of \sqrtsNNfive. 
The \Lcplus and \Lcminus baryons were reconstructed in the hadronic decay channels \LctopKpi and \LctopKzS and respective charge conjugates.
The measured differential cross sections as a function of transverse momentum ($\pt$) and the \pt-integrated $\Lcplus$ production cross section in \pp and in \pPb collisions are presented.
The \Lcplus nuclear modification factor (\RpPb), calculated from the cross sections in \pp and in \pPb collisions, is presented and compared with the \RpPb of D mesons.  
The \LcD ratio is also presented and compared with the light-flavour baryon-to-meson ratios p$/\pi$ and $\Lambda /\Kzs$, and measurements from other LHC experiments.
The results are compared to predictions from model calculations and Monte Carlo event generators.

\end{abstract}
\end{titlepage}

\setcounter{page}{2} 


\section{Introduction}
\label{sec: Introduction}
 
In hadronic collisions, heavy quarks
(charm and beauty) are created predominantly in hard scattering processes, and therefore the measurement of charm and beauty hadron production is a powerful test of perturbative quantum chromodynamics (pQCD) calculations.
Theoretical predictions based on the QCD factorisation approach describe the heavy-flavour hadron production cross section as a convolution of parton distribution functions, parton hard-scattering cross sections, and fragmentation functions. 
The measurements of D- and B-meson production cross sections in pp collisions at centre-of-mass energies between 200~GeV and 13~TeV at RHIC~\cite{Adamczyk:2012af}, Tevatron~\cite{Acosta:2003ax,Acosta:2004yw,Abulencia:2006ps}, and the LHC~\cite{Andronic:2015wma,Aaij:2015bpa,Khachatryan:2016csy,Aaij:2017qml,Acharya:2019mgn} are generally described within uncertainties by perturbative calculations at next-to-leading order with next-to-leading-log resummation, such as the general-mass variable-flavour-number scheme (GM-VFNS~\cite{Kniehl:2005mk,Kniehl:2012ti}) and fixed-order next-to-leading-log (FONLL~\cite{Cacciari:1998it,Cacciari:2012ny}), over a wide range of transverse momentum (\pT).
 
The measurement of the relative production of different heavy-flavour hadron species is also sensitive to the charm- and beauty-quark fragmentation and heavy-flavour hadron formation processes. 
In particular, measurements of the \Lc production cross section relative to D mesons provide insight into the hadronisation of charm quarks into baryons.
A measurement of \Lc baryon production at midrapidity in \pp collisions at \sqrtsseven was reported by the ALICE Collaboration in~\cite{Acharya:2017kfy}.
The \LcD ratio was found to be substantially higher than previous measurements at lower energies in electron-positron (\ee)~\cite{Albrecht:1988an,Avery:1990bc,Albrecht:1991ss,Gladilin:2014tba} and electron-proton (\ep)~\cite{Chekanov:2005mm, Abramowicz:2013eja, Abramowicz:2010aa} collisions, challenging the assumption that the probabilities for a charm quark to hadronise into a specific charm hadron (fragmentation fractions) are universal among different collision systems~\cite{Lisovyi:2015uqa}.   
In addition, the \LcD ratio was compared with predictions from several Monte Carlo (MC) generators, which implement different fragmentation processes, such as the formation of strings (PYTHIA\cite{Skands:2014pea,Christiansen:2015yqa}), ropes (DIPSY\cite{Bierlich:2015rha,Flensburg:2011kk}), or baryonic clusters (HERWIG\cite{Bellm:2015jjp}), where the fragmentation parameters for these simulations are tuned to previous \ee and \ep collision measurements.
These predictions significantly underestimate the \LcD ratio, although the prediction from \mbox{PYTHIA 8} that includes additional colour reconnection mechanisms~\cite{Christiansen:2015yqa} shows a \pt trend that is qualitatively similar to the measured trend.  
The CMS Collaboration has measured the \LcD ratio in pp collisions at \sqrtsfive~\cite{Sirunyan:2019fnc}, which is consistent with predictions from \mbox{PYTHIA 8} with additional colour reconnection mechanisms. 
\Lc production was also measured by the LHCb Collaboration in \pp collisions at \sqrtsseven at forward rapidity~\cite{Aaij:2013mga}, and the \LcD ratio was found to be lower than that measured by ALICE at midrapidity~\cite{Acharya:2017kfy}. 
Calculations of the charm-hadron production cross section based on the $k_\mathrm{T}$-factorisation approach with gluon distributions obtained on the basis of novel collinear gluon distribution functions and Peterson fragmentation functions~\cite{Maciula:2018iuh} are unable to simultaneously describe the ALICE and LHCb measurements using the same set of input parameters, suggesting that the measurements are difficult to explain within the independent parton fragmentation scheme. 
It is also important to note here that the magnitude of the relative production of $\Lbzero$ baryons and beauty mesons in pp collisions measured by LHCb~\cite{Aaij:2011jp,Aaij:2015fea,Aaij:2019pqz} and CMS~\cite{Chatrchyan:2012xg} offer further hints that fragmentation fractions in the beauty sector differ between pp and \ee/\ep collisions.

Measurements in pp collisions also provide a necessary reference for studies in heavy-ion collisions, where the study of charm production is a powerful tool to investigate the quark--gluon plasma (QGP)\cite{Adams:2005dq,Adcox:2004mh, Arsene:2004fa}, the deconfined state of matter created under extreme energy densities.   
In particular, the charm baryon-to-meson ratio in heavy-ion collisions is sensitive to the charm hadronisation mechanisms after the QGP phase. 
It is expected that a significant fraction of low- and intermediate-momentum charm 
quarks hadronise via recombination (coalescence) with light (anti) quarks from the medium\cite{Greco:2003vf,Oh:2009zj}, which would manifest as an enhancement of the \LcD ratio with respect to pp collisions. 
The \LcD ratio has been measured by STAR~\cite{Adam:2019hpq} in \AuAu collisions at \sqrtsNNtwohun, and by ALICE~\cite{Acharya:2018ckj} and CMS~\cite{Sirunyan:2019fnc} in \PbPb collisions at \sqrtsNNfive. These measurements offer constraints to different model calculations which implement contributions to hadronisation via quark recombination~\cite{Lee:2007wr, Zhao:2018jlw, Cho:2019lxb, He:2019vgs}.

The interpretation of the results obtained in heavy-ion collisions also requires detailed studies in \pPb collisions in order to assess so-called cold nuclear matter (CNM) effects in the initial and final states, which could modify the production of heavy-flavour hadrons. 
In the initial state, the quark and gluon distributions are modified in bound nucleons compared to free nucleons, depending on the fractional longitudinal parton momentum $x$ and the atomic mass number~\cite{ARNEODO1994301, Malace:2014uea}. The most relevant CNM effect at LHC energies is shadowing, i.e. a decrease of the parton densities in the small-$x$ region.
This effect is due to high phase-space densities of low-$x$ partons and can be described in collinear pQCD by means of parametrisations of the modification of the nuclear parton distribution functions (nPDFs)~\cite{Eskola:2016oht,Kovarik:2015cma}. 
In the case of saturation of the parton phase-space, the Colour Glass Condensate (CGC) effective theory~\cite{Gelis:2010nm,Tribedy:2011aa,Albacete:2012xq,Rezaeian:2012ye,Fujii:2013yja} offers an appropriate theoretical framework to describe the modification of the nPDFs.
Moreover, partons can lose energy in the initial stages of the collisions due to initial-state radiation~\cite{Vitev:2007ve}, or experience transverse momentum broadening due to multiple soft collisions before the heavy-quark pair is created in the hard scattering~\cite{Lev:1983hh,Wang:1998ww,Kopeliovich:2002yh}. The modification of parton distributions in the nucleus and energy loss in the initial state can affect the yields and the momentum distributions of the produced hadrons, mainly at low momenta.   
In addition to initial-state effects, 
final-state effects such as hadronic rescattering~\cite{Bierlich:2021poz} or the possible formation of a small QGP droplet~\cite{Beraudo:2015wsd,Xu:2015iha} can also modify the hadron yields and momentum distributions. Several measurements in high-multiplicity \pp and \pPb collisions, such as long-range correlations of charged hadrons~\cite{CMS:2012qk,Abelev:2012ola,ABELEV:2013wsa,Adam:2015bka}, and the enhancement of baryon-to-meson ratios in the light-flavour sector (p/$\pi$ and $\Lambda$/K)~\cite{Acharya:2018orn,Acharya:2020zji,Adam:2016dau}, exhibit a similar behaviour as that observed in \PbPb collisions, suggesting that these findings may have similar physical origins in pp, \pA, and A--A collisions~\cite{Nagle:2018nvi}.
\lambdac production was previously measured at midrapidity by ALICE in \pPb collisions at \sqrtsNNfive~\cite{Acharya:2017kfy}. The \LcD ratio was found to be compatible within the uncertainties with that measured in pp collisions at \sqrtsseven. The nuclear modification factor, \RpPb, was found to be compatible with unity, as well as with models that implement cold nuclear matter effects via nPDF calculations~\cite{Eskola:2009uj} or assume the production of a deconfined medium in \pPb collisions~\cite{Beraudo:2015wsd}. 
The LHCb Collaboration has measured the \LcD ratio at forward rapidity in \pPb collisions at $\sqrtsNN = 5.02~\tev$~\cite{Aaij:2018iyy} to be larger than that in pp collisions at forward rapidity~\cite{Aaij:2013mga} but smaller than the ALICE measurements in \pp and \pPb collisions at midrapidity~\cite{Acharya:2017kfy}.
 
Recent attempts have been made to model charm-baryon production in \pp and \pPb collisions. A framework based on a statistical hadronisation model~\cite{He:2019tik}, which takes into account an increased set of charm-baryon states beyond those listed by the Particle Data Group (PDG), is able to reproduce the \LcD ratios measured by ALICE in the \pp and \pPb collision systems, although it overestimates the LHCb measurement in pp collisions.
A model implementing hadronisation via recombination~\cite{Song:2018tpv,Li:2017zuj}, where the \pT distributions of light and charm quarks and antiquarks are inputs of the model and the relative production of single-charm baryons to single-charm mesons is treated as a free parameter, is able to reproduce the \pT dependence of the \LcD ratio measured by ALICE at central rapidity in pp and \pPb collisions, and by LHCb at forward rapidity in \pPb collisions. While models implementing different approaches to \Lc production are effective in describing
the measured \LcD ratio and \RpPb, the large statistical and systematic uncertainties of the current measurements do not provide the discriminating power needed to differentiate between the various models. Therefore, more precise measurements are crucial in order to constrain predictions.

This paper presents the measurement of the \pt-differential production cross section of charm \Lc baryons in \pp collisions in the rapidity interval $\absrap < 0.5$ and in \pPb collisions in $- 0.96 < y < 0.04$ at \sqrtsNNfive, performed with the ALICE detector at the LHC. The rapidity $y$ here and throughout this paper is defined in the centre-of-mass system, and in \pPb collisions the rapidity sign is positive in the p-going direction. The ratio of the production cross sections of \Lc baryons and \DZero mesons, \LcD, and the nuclear modification factor \RpPb are also presented. Finally, the \lambdac production cross section per unit of rapidity at midrapidity is computed by integrating the \pt-differential \lambdac production cross section after extrapolating down to \pT = 0, and the \pt-integrated \LcD ratios are presented.   
Two hadronic decay channels of \lambdac were studied: \LctopKpi and \LctopKzS. Different analysis strategies were implemented, taking advantage of the methods used in previous analyses for the hadronic decays of D mesons~\cite{Acharya:2017jgo,Adam:2016ich,Acharya:2018hre,ALICE:2011aa,Abelev:2012tca,Abelev:2012vra} and \lambdac baryons~\cite{Acharya:2017kfy}. 
With respect to our previous measurement of \Lc production~\cite{Acharya:2017kfy}, the \pT reach was extended, the overall uncertainties of the measurements were reduced, and the analysis was performed in finer \pT intervals.
The precision of the measurement of the nuclear modification factor \RpPb was improved with respect to the previously published result thanks to the larger data samples as well as a pp reference measured at the same centre-of-mass energy.  

The measurements are performed as the average of the particle and antiparticle cross sections, and so both \Lcplus and \Lcminus baryons are referred to collectively as \Lc in the following. 
In all measurements the production cross section of prompt \Lc is reported, i.e. \Lc from direct hadronisation of a charm quark or from decays of directly produced excited charm states. For the centre-of-mass energy of \pp collisions the simplified notation \sqrts is used throughout this paper. 

It is noted that the \LcD baryon-to-meson ratio is the focus of a dedicated letter~\cite{ALICE:2020wfu}, and this document presents a more detailed description of the analysis procedure as well as supplementary results.

\section{Experimental setup and data samples}
\label{sec: Data samples and experiment}
The ALICE 
apparatus is composed of a central barrel, consisting of a set of detectors for particle reconstruction and identification covering the midrapidity region, a muon spectrometer at forward rapidity and various forward and backward detectors for triggering and event characterisation. The central barrel detectors cover the full azimuth in the pseudorapidity interval $|\eta|<0.9$ and are
embedded in a large solenoidal magnet that provides a $B = 0.5$~T field parallel to the beam direction ($z$-axis in the ALICE reference frame).
A comprehensive description and overview of the typical performance of the detectors in pp and \pPb collisions
can be found in~\cite{Aamodt:2008zz, Abelev:2014ffa}.

The  tracking  and  particle  identification capabilities  of the ALICE central barrel detectors were exploited
to reconstruct the \Lc decay products at midrapidity. 
The Inner Tracking System (ITS), consisting of three subdetectors, the Silicon Pixel Detector (SPD),
the Silicon Drift Detector (SDD), and the Silicon Strip Detector (SSD), each made of two concentric layers, allows for a precise determination of the track impact parameter (the distance of closest approach between the track and the primary vertex of the collision) 
in the transverse plane with a resolution better than 75\,$\mu$m for tracks with $\pt > 1$~\gevc~\cite{Aamodt:2010aa}.
The Time Projection Chamber (TPC) is the main tracking detector of the experiment~\cite{Alme:2010ke}.
It provides up to 159 space points to reconstruct the charged-particle trajectory,
and provides charged-particle identification (PID) via the measurement of the specific energy loss \dEdx.
The particle identification capabilities are extended by the Time-of-Flight (TOF) detector,
which is used to measure the flight time of charged particles from the interaction point. 
The TOF detector is an array of Multi-gap Resistive Plate Chambers. 
It measures the particle arrival time at the detector with a resolution of about 80 ps. 
The start time of the collision is obtained for each event either using the TOF detector,
the T0 detector, or a combination of the two~\cite{Adam:2016ilk}.
The T0 detector consists of two arrays of Cherenkov counters, located on both sides of the interaction point, covering the pseudorapidity regions ${4.61<\eta<4.92}$ and ${-3.28 < \eta < -2.97}$, respectively. 
The time resolution of the T0 detector in \pp and \pPb collisions is about 50 ps for events
in which a measurement is made on both sides of the interaction point~\cite{Adam:2016ilk}.
The V0 detector system, used for triggering and event selection, consists of two scintillator arrays covering the full azimuth in the pseudorapidity intervals ${2.8 < \eta < 5.1}$ and ${-3.7 <\eta< -1.7}$ (\cite{Aamodt:2008zz}, section 5.1).  
The Zero Degree Calorimeter (ZDC), used for offline event rejection in \pPb collisions, consists of two sets of neutron and proton calorimeters positioned along the beam axis on both sides of the ALICE apparatus, about 110 m from the interaction point (\cite{Aamodt:2008zz}, section 5.4).

The results presented in this paper were obtained from the analysis of the LHC Run 2 data samples collected from \pp collisions 
at \sqrtsfive in 2017 and \pPb collisions at \sqrtsNNfive in 2016. 
The proton--nucleon centre-of-mass system in \pPb collisions is shifted in rapidity by $\Delta y$ = 0.465 in the Pb-going direction (negative rapidity)
due to the asymmetric beam energies of 4\,TeV for protons and 1.59\,TeV per nucleon for Pb nuclei.
The analyses used events recorded with a minimum bias (MB) trigger, which was based on coincident signals from the V0 detectors in both \pp and \pPb collisions. 
In order to remove background from beam--gas collisions and other machine-induced backgrounds, in \pp collisions the events were further selected offline based on the correlation between the numbers of clusters and track segments reconstructed in the SPD, and V0 timing information. The latter was also used for the \pPb analysis, together with the timing from the ZDC.
In order to maintain a uniform ITS acceptance in pseudorapidity, only events with a $z$-coordinate of the reconstructed vertex position within 10 cm from the nominal interaction point were analysed.
Events with multiple interaction vertices due to pileup from several collisions were removed using an algorithm based on tracks reconstructed with the TPC and ITS detectors~\cite{Abelev:2014ffa}. 
Using these selection criteria, approximately one billion MB-triggered \pp events were analysed,
corresponding to an integrated luminosity of $\lumi$ = 19.5 \nbinv ($\pm$2.1\%~\cite{ALICEpb2018}),
while approximately 600 million MB-triggered \pPb events were selected, corresponding to $\lumi = 287~\ubinv$ ($\pm$3.7\%~\cite{Abelev:2014epa}).
 
\section{\Lc analysis overview and methods}
\label{sec: Analysis overview and methods}

The analysis was performed using similar techniques to those reported in~\cite{Acharya:2017kfy}. \Lc baryons were reconstructed in two hadronic decay channels: \LcpKpi (branching ratio, BR $= 6.28\pm0.33\%$), and \LcpKs (BR $= 1.59\pm0.08\%$), followed by the subsequent decay $\Kzs\to\piPlus\piMinus$ (BR $= 69.2\pm0.05\%$)~\cite{PDG20}. For the former, the $\Lc$ decays to the $\rm pK^-\pi^+$ final state via four channels: $\Lc\to\mathrm{p\overline{K}^{*0}(892)}$, $\Lc\to\Delta^{++}(1232)\mathrm{K}^{-}$, $\Lc\to\Lambda(1520)\pi^{+}$, and the non-resonant \LctopKpi decay. As these channels are indistinguishable in the analysis, all four are considered together.

The selection of candidates was performed using a combination of kinematical, geometrical, and PID selections. 
The selection criteria were tuned on Monte Carlo simulations in order to maximise the statistical significance in each \pt interval. 
\Lc candidates were reconstructed by combining reconstructed tracks with $\abs{\eta} < 0.8$ and at least 70 reconstructed space points in the TPC. For all decay products in the \LctopKpi analysis and for the proton-candidate tracks in the \LctopKzS analysis, at least one cluster was required in either of the two SPD layers. The PID selections for all analyses were performed utilising the Bayesian method for combining the TPC and TOF signals, as described in~\cite{Adam:2016acv}. 
The Bayesian method entails the use of priors, an \textit{a priori} probabilitiy of measuring a given particle species, which are determined using measured particle abundances.
Where possible, the TPC and TOF signals were combined; however, if the TOF signal was absent for a given track, the TPC signal alone was used.
For the \LcpKs analysis in \pPb collisions, a machine learning approach with Boosted Decision Trees (BDTs) was applied to select \Lc candidates, using the Toolkit for Multivariate Data Analysis (TMVA)~\cite{Hocker:2007ht}.

The detector acceptance for \Lc baryons varies as a function of rapidity, in particular falling steeply to zero for $\absrap > 0.5$ at low \pt, and $\absrap > 0.8$ for $\pt>5\,\GeVc$. For this reason, a fiducial acceptance selection was applied on the rapidity of candidates, $\abs{y_\mathrm{lab}} < y_\mathrm{fid}(\pt)$, where $y_\mathrm{fid}$ increases smoothly from 0.5 to 0.8 in $0<\pt<5~\GeVc$ and $y_\mathrm{fid}=0.8$ for $\pt>5\,\GeVc$~\cite{Acharya:2017jgo}.

For the \LcpKpi analysis, candidates were formed by combining triplets of tracks with the correct configuration of charge sign. For this decay channel, the high-resolution tracking and vertexing information provided by the ITS and TPC allows the interaction point (primary vertex) and the reconstructed decay point of the \Lc candidate (secondary vertex) to be distinguished from one another, despite the short decay length of the \Lc ($\ctau=60.7\,\mim$~\cite{PDG20}).
Once the secondary vertex was computed from the three tracks forming the \Lc candidate, selections were applied on variables related to the 
kinematic properties of the decay, the quality of the reconstructed vertex, and the displaced decay-vertex topology.
These variables comprise the transverse momenta of the decay products; the quadratic sum of the distance of closest approach of each track to the
secondary vertex; the decay length of the \Lc candidate (separation between the primary and
secondary vertices); and the cosine of the pointing angle between the \Lc candidate flight line (the
vector that connects the primary and secondary vertices) and the reconstructed momentum
vector of the candidate.
Pions, kaons, and protons were identified using the \textit{maximum-probability} Bayesian PID approach~\cite{Adam:2016acv}, where a probability is assigned to each track for every possible species based on the TPC and TOF signals and the identity of the track is taken to be the species with the highest probability value.
This approach allows for a higher-purity sample to be selected, reducing the large level of combinatorial background and facilitating the signal extraction.

The \LcpKs analysis started from a $\Kzs\to\piPlus\piMinus$ candidate, which is reconstructed as a pair of opposite-sign charged tracks forming a neutral decay vertex displaced from the primary vertex (a \Vzero candidate). This \Vzero candidate was paired with a proton-candidate track originating from the primary vertex to form a \Lc candidate. Two strategies were then used to select \Lc candidates in \pp and \pPb collisions. In \pp collisions, the analysis was based on rectangular selection criteria. 
The \Vzero candidate was required to have an invariant mass compatible with the \Kzs mass from the PDG~\cite{PDG20} within $8~(20)~\mevcc$ at low (high) $\pt$, corresponding to one or two times the resolution of the \Kzs invariant mass, depending on the \pt interval and the collision system. 
The \Vzero candidates were selected based on the \pT and impact parameter of the decay pions to the \Kzs decay vertex, and the cosine of the pointing angle between the \Vzero flight line and its reconstructed momentum. Proton-candidate tracks were selected based on their \pT, their impact parameter to the primary vertex, the number of reconstructed TPC clusters, and a cluster being present on at least one of the two SPD layers. Particle identification was performed on the proton-candidate track, first using a loose $\abs{\nsigma} < 3$ pre-selection on the TPC response, where $\nsigma$ corresponds to the difference between the measured and expected \dEdx for a given particle species, in units of the resolution. This was followed by a strict requirement that the Bayesian posterior probability for the track to be a proton must be greater than 80\%.

In \pPb collisions, an approach using BDTs was used for the \LcpKs decay. The BDT algorithm provides a classification tree that maps simulated $\Lc$ candidates to a single BDT response variable aiming to maximise the separation between signal and background candidates. The mapping function is then applied on a real data sample in which the true identities of particles are unknown, followed by the application of selections on the BDT response. 
Candidates were initially filtered using an $|\nsigma^\mathrm{TPC}| <3$ PID selection on the proton candidate. Independent BDTs were trained for each \pt interval in the analysis. The training
was performed on samples of simulated events including a detailed description of the experimental apparatus and the detector response. 
The training sample for signal candidates was taken from a simulation of pp events containing charm hadrons generated using \mbox{PYTHIA 6.4.25}~\cite{Sjostrand:2006za} with the Perugia2011 tune~\cite{Skands:2009zm}, embedded into an underlying \pPb collision generated with HIJING 1.36~\cite{Wang:1991hta}.
The background candidates were taken from the HIJING simulation.
The variables that were used in the training were the Bayesian PID probability of the proton-candidate track to be a proton, the \pt of the proton candidate, the invariant mass and $c\tau$ of the \Kzs candidate, and the impact parameters of the \Vzero and the proton-candidate track with respect to the primary vertex. 
The MC samples used for the efficiency calculation were different from those used in the training. 
The selection on the BDT response was tuned in each \pt interval to maximise the expected statistical significance, which is estimated using i) the signal obtained from the generated \Lc yield multiplied by the selection efficiency of the trained model and ii) the background estimated from preselected data multiplied by the background rejection factor from the BDT.
The BDT analysis was cross checked with an independent analysis using rectangular selection criteria, and the two results were found to be fully consistent within the experimental uncertainties.

\begin{figure}[t!b]\centering
\includegraphics[width=0.4\textwidth]{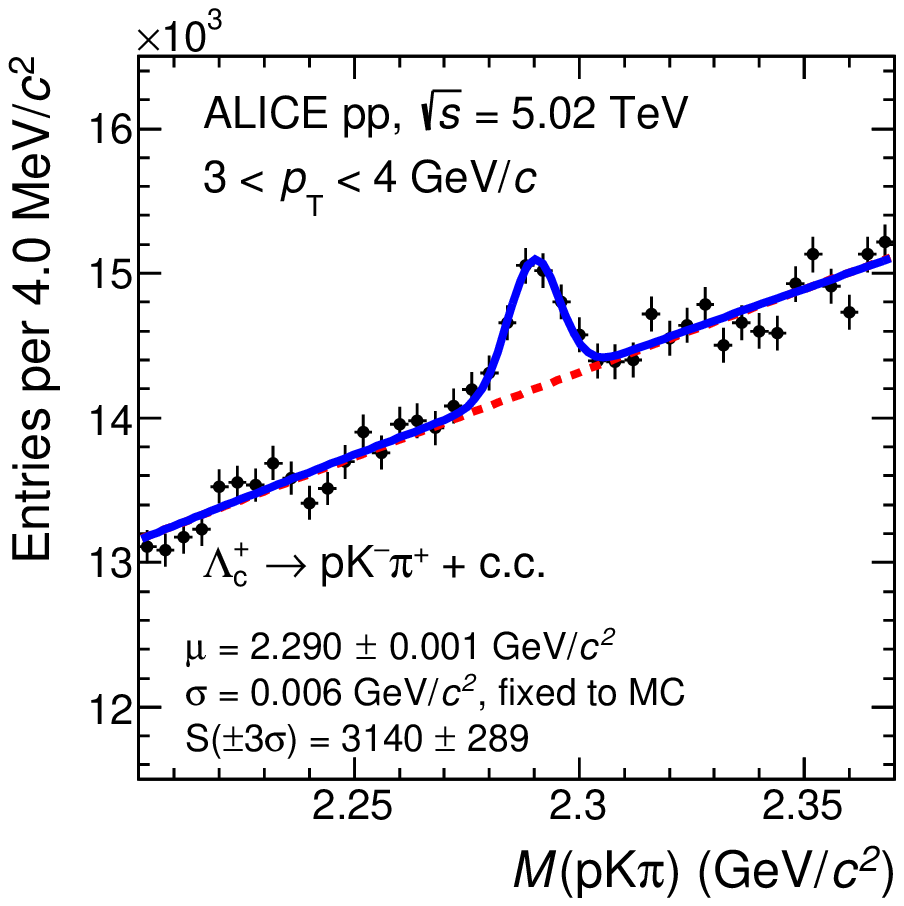}
\includegraphics[width=0.4\textwidth]{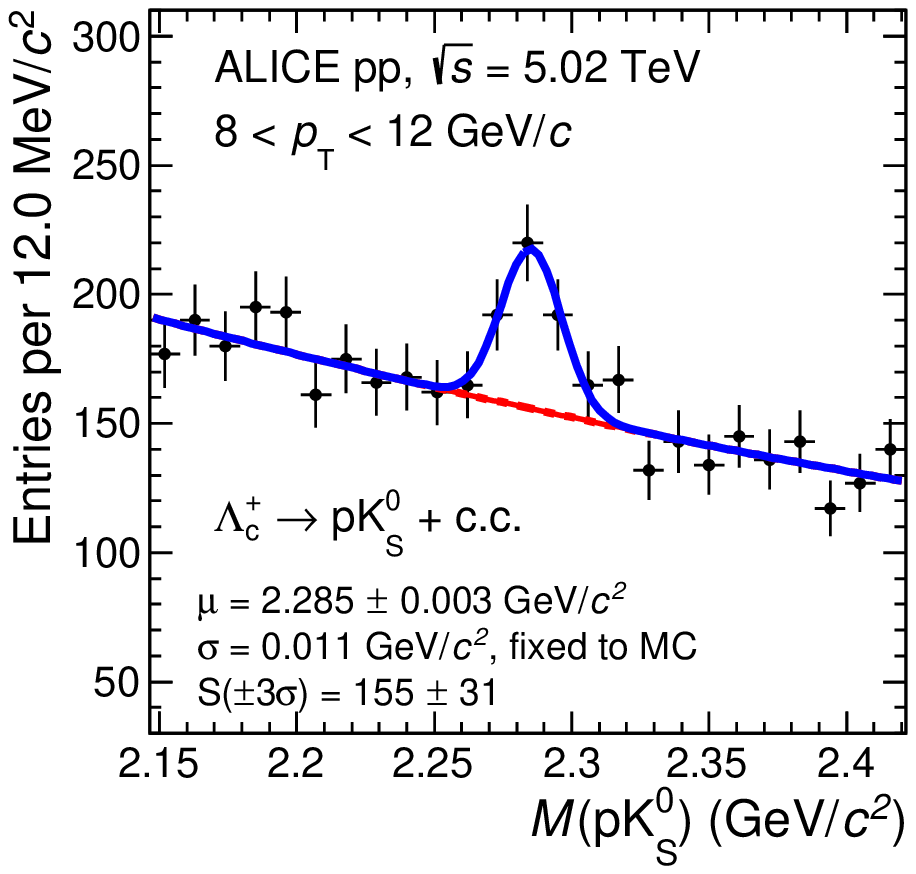}\\
\includegraphics[width=0.4\textwidth]{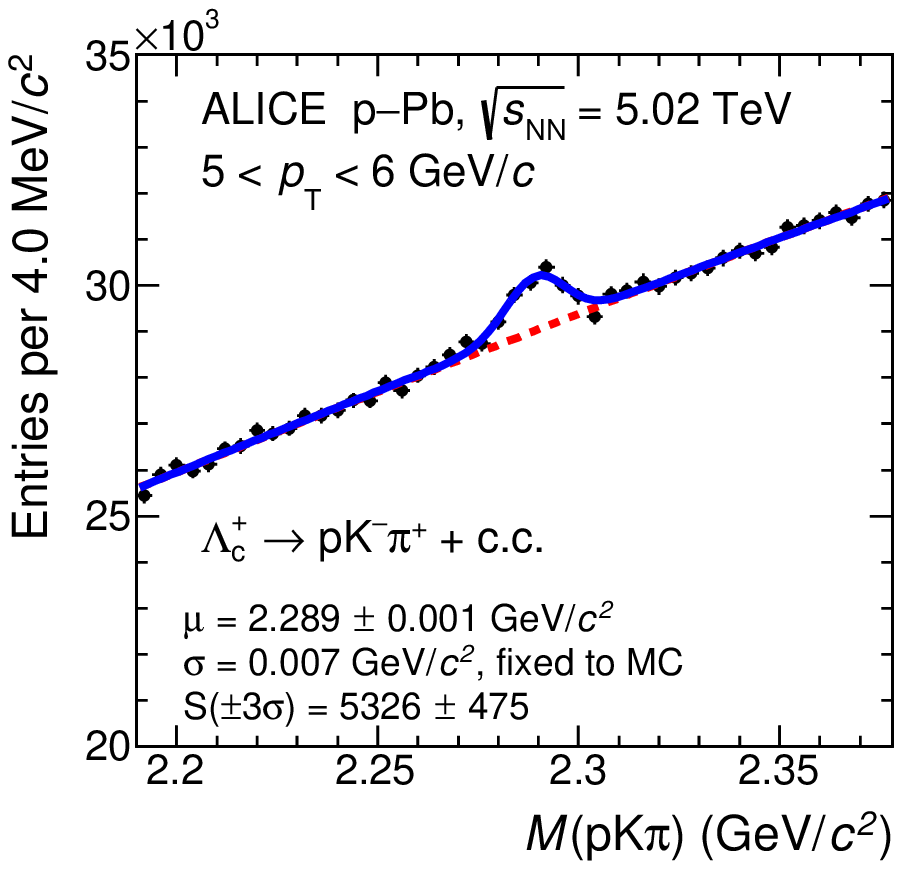}
\includegraphics[width=0.4\textwidth]{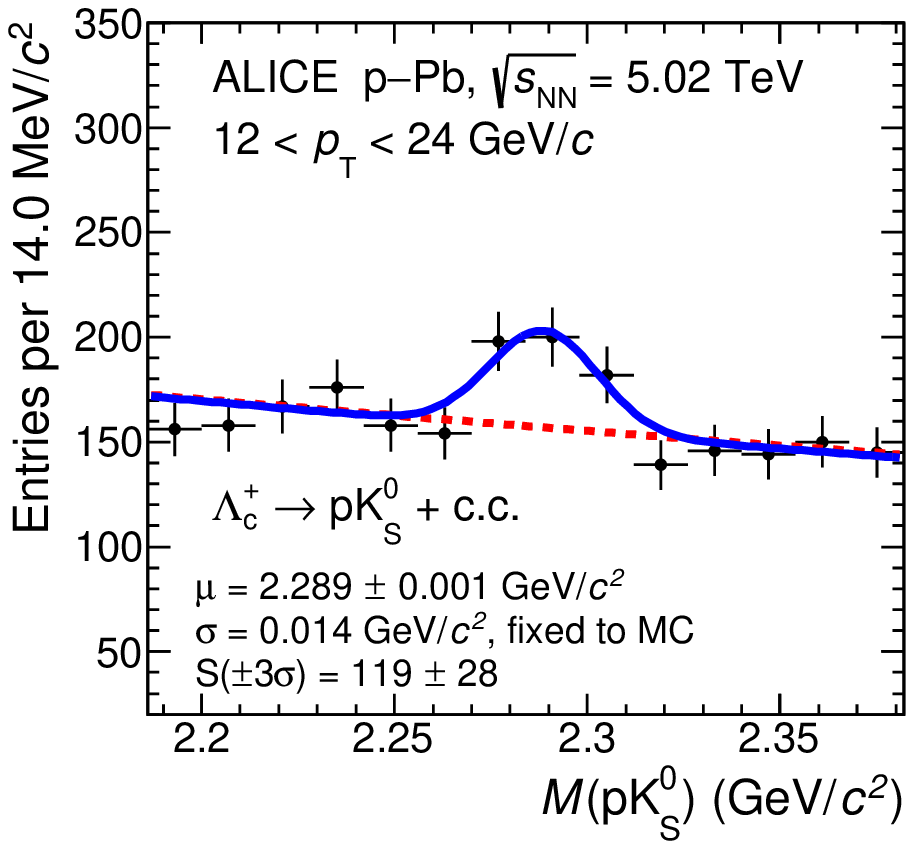}
\caption{Invariant mass distributions of \Lc candidates in different \pt intervals, collision systems, and decay channels, with the corresponding fit functions. Top-left: \LcpKpi for $3<\pt<4\,~\GeVc$ in pp collisions; top-right: \LcpKs for $8<\pt<12\,~\GeVc$ in pp collisions; bottom-left: \LcpKpi for $5<\pt<6\,~\GeVc$ in \pPb collisions; bottom-right: \LcpKs with BDT analysis in $12<\pt<24\,~\GeVc$ in \pPb collisions. The dashed lines represent the fit to the background and the solid lines represent the total fit function.}
\label{fig:analysis_invmass}
\end{figure}

Signal extraction for all analyses was performed by means of a fit to the invariant mass distributions of candidates in each \pT interval under study. A Gaussian function was used to model the signal peak and an exponential or polynomial function was used to model the background. Due to the small signal-to-background ratio, the standard deviation of the Gaussian signal function was fixed to the value obtained from simulations in order to improve the fit stability. In \pp collisions, a \Lc signal could be extracted for the $\LctopKpi$ and $\LctopKzS$ analyses in the range $1 < \pt < 12~\gev$. In \pPb collisions a \Lc signal was extracted for the $\LctopKzS$ analysis in the range $1 < \pt < 24~\gevc$, and for the $\LctopKpi$ analysis in the range $2 < \pt < 24~\gevc$, as the larger combinatorial background in the $\LctopKpi$ channel limits the low-\pt reach.
A selection of the invariant mass distributions with their corresponding fit functions is displayed in \figref{fig:analysis_invmass} for different \pt intervals, decay channels, and collision systems.

\section{Corrections}
\label{sec: Corrections}

The \pt-differential cross section of prompt \Lcplus-baryon production was obtained for each decay channel as

\begin{equation} 
\label{eqn:lc_normalisation}
\frac{\mathrm{d^2}\sigma^{\Lambda_{\rm c}^+}}{\mathrm{d}\pt\mathrm{d}y} = \frac{1}{2 c_{\Delta y} \times \Delta \pt} \times \frac{1}{\rm{BR}} \times
\frac{  f_{\rm prompt} \times N^{\Lambda_{\rm c}}_{\lvert y \lvert < y_{\rm fid}}}{ (A\times\varepsilon)_{\rm prompt}} \times \frac{1}{\lumi},
\end{equation}

where $N^{\Lambda_{\rm c}}$ is the raw yield (sum of particles and antiparticles) in a given $\pt$ interval with width $\Delta \pt$, $f_{\rm prompt}$ is the fraction of the raw yield from prompt \Lc, BR is the branching ratio for the considered decay mode, and $\lumi$ is the integrated luminosity. $(A\times\varepsilon)$ is the product of detector acceptance and efficiency for prompt \Lc baryons, where $\epsilon$ accounts for the reconstruction of the collision vertex, the reconstruction and selection of the tracks of the \Lc decay products, and the \Lc-candidate selection.
The correction factor for the rapidity coverage, $c_{\Delta y}$, was computed as the ratio between the generated \Lc-baryon yield in 
$|y_{\rm lab}|<y_{\rm fid}$(\pt) and that in $|y_{\rm lab}|<0.5$, where the \Lc-baryon rapidity shape was taken from FONLL pQCD calculations.
The factor 2 in the denominator of Eq.~\ref{eqn:lc_normalisation} takes into account that the raw yield includes both particles and antiparticles, while the cross section is given for particles only and is computed as the 
average of \Lcplus~and \Lcminus.

The correction factor $(A \times\varepsilon)$ was 
obtained following the same approach as discussed in~\cite{ALICE:2011aa}. 
The correction factors were obtained from simulations in which the detector
and data taking conditions of the corresponding data samples were reproduced.
\mbox{PYTHIA 6.4.25} and \mbox{PYTHIA 8.243}~\cite{Sjostrand:2007gs} were used to simulate pp collisions.
For \pPb collisions, a pp event containing heavy-flavour signals was generated with \mbox{PYTHIA 6} and HIJING was used to simulate the underlying background event.

The $(A\times\varepsilon)$ was computed separately for prompt and non-prompt \Lc. 
The \LctopKpi decay channel includes not only the direct (non-resonant) decay mode, but also three resonant channels, as explained in Section~\ref{sec: Analysis overview and methods}. 
Due to the kinematical properties of these decays, the acceptance and efficiency of each decay mode is different
and the final correction was determined as a weighted average of the $(A\times\varepsilon)$ values of the four decay channels with the relative branching ratios as weights.

\begin{figure}[t!]
\centering
\includegraphics[width=0.7\textwidth]{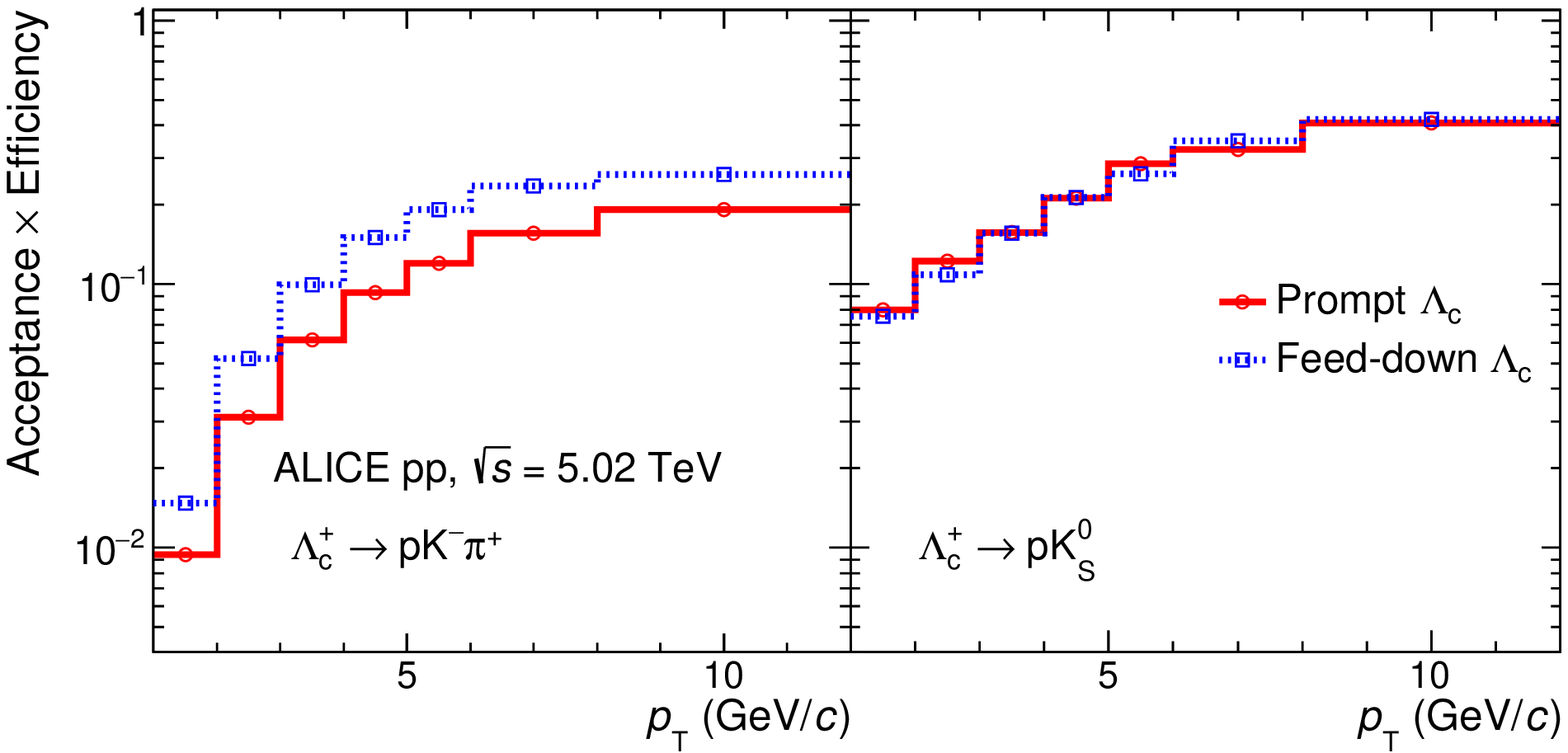}
\caption{Product of detector acceptance and efficiency 
for \Lc baryons in pp collisions at $\sqrt{s}=5.02~\TeV$,
  as a function of \pt. From left to right: \LctopKpi and \LctopKzS. The solid lines correspond to the $(A\times\varepsilon)$ for prompt \Lc, while the dotted lines represent $(A \times \epsilon)$ for \Lc baryons originating from beauty-hadron decays. The statistical uncertainties are smaller than the marker size. }
\label{fig:AccEff_pp5TeV}
\end{figure}

\begin{figure}[t!]
\centering
\includegraphics[width=0.7\textwidth]{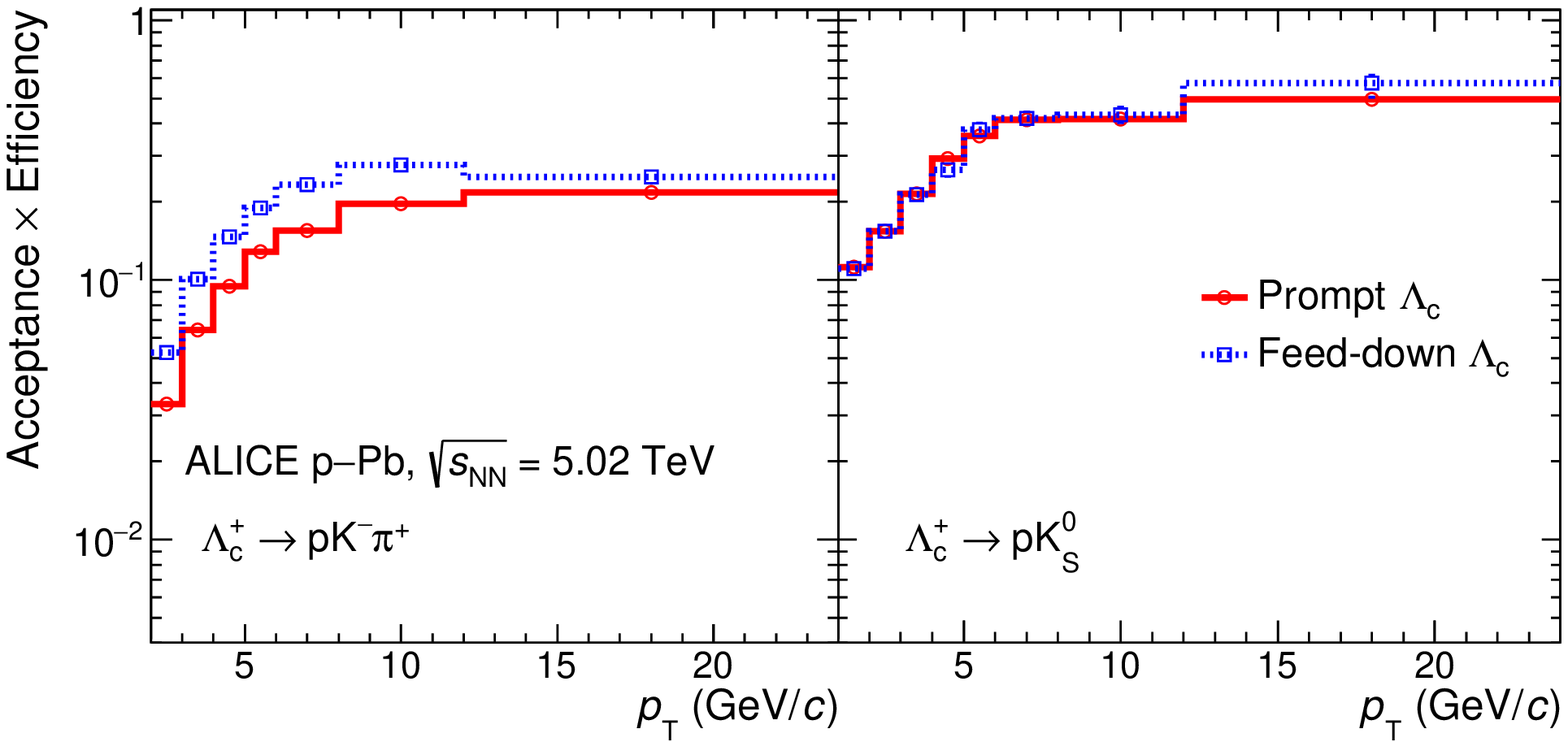}
\caption{Product of detector acceptance and efficiency 
for \Lc baryons in \pPb collisions at \sqrtsNNfive, as a function of \pt. From left to right: \LctopKpi and \LctopKzS. The solid lines correspond to the $(A\times\varepsilon)$ for prompt \Lc, while the dotted lines represent $(A \times \epsilon)$ for \Lc baryons originating from beauty-hadron decays. The statistical uncertainties are smaller than the marker size. }
\label{fig:AccEff_pPb5TeV}
\end{figure}

\Figrefs{fig:AccEff_pp5TeV}{fig:AccEff_pPb5TeV} show the product of $(A\times\varepsilon)$ for \Lc
baryons with $|y| < y_{\rm fid}$ in pp and \pPb collisions
as a function of \pt for the \LctopKpi (left panel) and \LctopKzS (right panel) decay channels.
The higher $(A\times\varepsilon)$ for \Lc from beauty-hadron decays in the \LctopKpi decay channel is due to the geometrical selections on the displaced decay-vertex topology,
which enhance the non-prompt component because of the relatively longer lifetime of the beauty hadrons compared to prompt \Lc.
For the \LctopKzS analyses, the $(A\times\varepsilon)$ of prompt and non-prompt \Lc are compatible, as selections based on the displaced decay-vertex topology are not applied.

Contrary to \pp collisions, where the charged-particle multiplicity in data is well described by the simulation, in \pPb collisions a weighting procedure based on the
event multiplicity was used in the calculation of the reconstruction efficiency from the simulated events. This approach accounts for the dependence of the reconstruction efficiency on the event multiplicity, which is due to
the fact that the resolutions of the primary-vertex position and of the variables used in the geometrical selections of displaced decay vertices improve with increasing multiplicity. The event multiplicity was defined here using the number of tracklets, where a tracklet is defined as a track segment joining the reconstructed primary vertex with a space point on each SPD layer within the pseudorapidity range $|\eta| < 1.0$.

The factor $f_{\textrm{prompt}}$ was calculated as in~\cite{Acharya:2017kfy}:

\begin{equation}	
f_{\textrm{prompt}} = 1 - \frac{N^{\Lambda_c\text{feed-down}}}{N^{\Lambda_c}} = 1 - \frac{(A \times \epsilon)_{\text{feed-down}}~c_{\Delta y}~\Delta p_{\textrm{T}}~{\rm{BR}}~{\lumi}}{N^{\Lambda_c}/2} \times \left(\frac{\textrm{d}^{2}\sigma}{\textrm{d}p_{\textrm{T}}\textrm{d}y}\right)^{\rm FONLL}_{\text{feed-down}},
\label{eq:nb}
\end{equation}

where $N^{\Lambda_c}/2$ is the raw yield divided by a factor of two to account for particles and antiparticles.
The production cross section of \Lc from beauty-hadron decays, $\left(\frac{\textrm{d}^{2}\sigma}{\textrm{d}p_{\textrm{T}}\textrm{d}y}\right)^{\rm FONLL}_{\rm feed-down}$, was calculated using the b-quark \pt-differential cross section from FONLL calculations~\cite{Cacciari:1998it,Cacciari:2012ny}, the fraction of beauty quarks that fragment into beauty hadrons $\mathrm{{H}_{b}}$ estimated from LHCb measurements~\cite{Aaij:2019pqz}, and the $\mathrm{{H}_{b}}$ $\rightarrow$ \Lc + {\it{X}} decay kinematics and branching ratios of $f(\mathrm{{H}_{b}} \rightarrow \Lc + X)$ modelled using \mbox{PYTHIA 8} simulations~\cite{Sjostrand:2007gs}.

The beauty-hadron fragmentation was derived from the LHCb measurements of the \Bsbar- and \Lb-production fraction relative to \BZerobar and \Bminus mesons in pp collisions at $\sqrt{s} = 13~\tev$~\cite{Aaij:2019pqz}, which indicates that the fraction of b quarks hadronising into a \Lb baryon is strongly $\pt$-dependent in the measured range of $4 < \pt < 25~\gevc$.
The fits to the production fractions of $\Bsbar$ and $\Lb$ hadrons normalised to the sum of $\Bminus$ and $\BZerobar$ hadrons are presented in~\cite{Aaij:2019pqz} as a function of the beauty-hadron \pt as

\begin{equation} 
	\frac{f_{s}}{f_u + f_d} (\pt) = A[p_1 + p_2 \times (\pt - <\pt>)] = X , 
\end{equation}
\begin{equation} 
	\frac{f_{\Lb}}{f_u + f_d} (\pt) = C[q_1 + \exp(q_2 + q_3 \times \pt)] = Y ,
\end{equation}

where $f_u$, $f_d$, $f_s$, and $f_{\Lb}$ are the fractions of b quarks that hadronise into $\BZerobar$, $\Bminus$, $\Bsbar$, and $\Lb$, respectively, and $A$, $p_1$, $p_2$, $<\pt>$, $C$, $q_1$, $q_2$ and $q_3$ are free parameters of the fits to the measured ratios. The beauty hadron fragmentation fractions are defined assuming $f_u = f_d$ and $f_u+f_d+f_s+f_{\Lb} = 1$. Around 90\% of the feed-down \Lc comes from $\Lb$ $\rightarrow$ \Lc + {\it{X}} decays, and the $\Lb$ fragmentation fraction can be defined as

\begin{equation} 
	f_{\Lb} (\pt) = \frac{Y}{(X + Y + 1)} .
\end{equation}

For $\pt=5~\gevc$,
$f_{\Lb}$ is around 0.2,
and it decreases to a value of around 0.09 for $\pt>20~\gevc$. For $\pt<5~\gevc$ it was assumed that $f_{\Lb}=0.2$, since measurements of the ratio $\Lb / \BZerobar$ in pp collisions at $\sqrt{s} = 7$ TeV and 8 TeV~\cite{Aaij:2015fea} are flat as a function of \pt in this interval within the experimental uncertainties. It was assumed that there is no rapidity dependence of $f_{\Lb}$ since the LHCb measurements of beauty-production ratios are flat as a function of rapidity in $2 < y < 5$ within the experimental uncertainties~\cite{Aaij:2015fea,Aaij:2019pqz}. 

For \pPb collisions, a hypothesis on the nuclear modification factor $R_{\textrm{pPb}}^{\textrm{feed-down}}$ of \Lc from beauty-hadron decays
was included as an additional factor in the last term of Eq.~\ref{eq:nb}. As in the D-meson analyses~\cite{Adam:2016ich},
it was assumed that the $\RpPb$ of prompt and feed-down \Lc are equal. 
The values of $f_{\textrm{prompt}}$ in both collision systems range between 87\% and 98\% for the \LcpKs decay channel and between 84\% and 98\% for the \LcpKpi decay channel.

\section{Evaluation of systematic uncertainties}
\label{sec: Systematics}

This section describes the various sources of systematic uncertainties of the measured cross section in each analysis, and the methods used to estimate them. A summary of the systematic uncertainties is shown in \tabref{tab:Syspp} and \tabref{tab:SyspPb} for the \pp and \pPb analyses, respectively. The different sources of systematic
uncertainty are assumed to be uncorrelated, and their contributions are added in quadrature to calculate the overall systematic uncertainty in each \pt interval.

The systematic uncertainty on the yield extraction was estimated by repeating the fits to the invariant mass distributions several times, varying i) the lower and upper limits of the fit interval, and ii) the functional form of the background (linear, exponential, and second-order polynomial functions were used). For each of the above trials, the fit was repeated with different hypotheses on the signal peak width and mean, with variations including a) treating both the Gaussian width and mean as free parameters, b) fixing the peak width to the MC expectation and leaving the mean free, c) fixing the mean to the MC expectation and leaving the peak width free, and d) fixing both the peak width and mean to the MC expectation. The systematic uncertainty was defined as the RMS of the distribution of the raw yield values extracted from these trials.

The systematic uncertainty on the tracking efficiency was estimated by i) comparing the probability of prolonging a track from the TPC to
the ITS (``matching efficiency'') in data and simulation, and ii) by varying track selection criteria in the analyses. The matching efficiency in simulation was determined after re-weighting the relative abundance of primary and secondary particles to match that in data. 
The uncertainty on the matching efficiency was defined as the relative difference in the matching efficiency between simulation and data. It is species-dependent and therefore it was determined individually for protons, kaons, and pions. In the \LctopKzS analysis only the proton matching efficiency uncertainty was included since no ITS condition was required for the pion tracks from the \Kzs decay. 
The per-track uncertainty on the matching efficiency is \pt dependent and it was propagated to the \Lc taking into account the decay kinematics and treating the uncertainty as correlated among the tracks. The second contribution to the track reconstruction uncertainty was estimated by repeating the analysis varying the TPC track selection criteria. The uncertainty was defined as the RMS of the \Lc cross section values obtained with the different track selections. The total uncertainty on the tracking efficiency was defined as the quadratic sum of these two contributions. 

The uncertainty on the \Lc selection efficiency due to imperfections in the simulated kinematical and geometrical variables used to select \Lc candidates was estimated by varying the selection criteria. 
For the BDT analysis in the \LctopKzS channel, variations were made on the selection of the BDT response.
The systematic uncertainty was estimated in each \pt interval as the RMS of the distribution of the corrected cross section values resulting from these variations.

Systematic uncertainties can arise from discrepancies in the PID efficiency between simulation and data. In the case of the \LctopKzS analysis in \pp collisions, the systematic uncertainty associated with the PID efficiency was estimated by varying the minimum probability threshold required to identify a track as a proton. For the \LctopKpi analysis, the systematic uncertainty was estimated by applying a minimum threshold selection on the Bayesian probability to assign the track identity, with the threshold varying between 30\% and 80\%. The systematic uncertainty in both cases was defined based on the variation of the corrected cross section. For the \LctopKzS analysis in \pPb collisions, the PID variables were included as part of the BDT, and therefore the PID uncertainty is already accounted for by varying the selection on the BDT response. The contribution due to the $3\sigma$ PID preselection was found to be negligible.

An additional source of systematic uncertainty was assigned due to the dependence of the efficiencies on the generated \pt distribution of \Lc in the simulation (``MC $\pt$ shape'' in \tabref{tab:Syspp} and~\ref{tab:SyspPb}). To estimate this effect the efficiencies were evaluated after reweighting the \pt shape of the PYTHIA 6 simulations to match the \pt spectrum of D mesons from FONLL pQCD calculations. An uncertainty was assigned in each \pt interval based on the difference between the central and reweighted efficiencies.

The relative statistical uncertainty on $(A\times\varepsilon)$ was considered as an additional systematic uncertainty source, originating from the finite statistics in the simulation used to calculate the efficiency. 

The systematic uncertainty on the prompt fraction (``Beauty feed-down'' in \tabref{tab:Syspp} and \ref{tab:SyspPb}) was estimated by varying independently i) the production cross section of beauty quarks within the theoretical uncertainties in FONLL~\cite{Cacciari:2012ny}, and ii) the function describing the fragmentation fraction $f_{\Lb}$. For the variation of ii), the free parameters defined in~\cite{Aaij:2019pqz} were varied independently within their uncertainties. For $\pt(\Lb)<5~\gevc$, the lower uncertainty bound of $f_{\Lb}$ was taken to be equal to the lower bound of the fit at $\pt(\Lb)=5~\gevc$, independent of \pt, while the upper uncertainty bound was taken to be equal to the \pt-dependent upper bound of the fit. In order to account for a possible \sqrts dependence of the fragmentation fractions, an additional reduction of the lower bound of $f_{\Lb}$ was considered based on the spread of the LHCb measurements at different values of \sqrts. In the \pPb analyses the uncertainty on the hypothesis of the nuclear modification factor of \Lc from beauty-hadron decays was estimated by varying the ratio $R_{\textrm{pPb}}^{\textrm{feed-down}}/R_{\textrm{pPb}}^{\textrm{prompt}}$ in the range $ 0.9 < R_{\textrm{pPb}}^{\textrm{feed-down}}/R_{\textrm{pPb}}^{\textrm{prompt}} < 1.3$. This range was chosen based on theoretical calculations of charm and beauty hadron production in \pPb collisions as explained in~\cite{Adam:2016ich}. The overall uncertainty on the prompt fraction was defined as the envelope of these variations, which leads to an asymmetric uncertainty.

The uncertainty on the luminosity measurement is 2.1\% for pp collisions~\cite{ALICEpb2018} and 3.7\% for \pPb collisions~\cite{Abelev:2014epa}. The uncertainty on the branching fractions are 5.1\% for the \LctopKpi channel, and 5.0\% for the \LctopKzS channel~\cite{PDG20}.

\begin{table}[!htb]
	
	\begin{center}
		\def\arraystretch{1.2}\tabcolsep=2pt    
		\begin{tabular}{lccccc} 
			\toprule
			& \multicolumn{2}{c}{$\LctopKpi$} & & \multicolumn{2}{c}{$\LctopKzS$} \\
			\cmidrule{2-3} \cmidrule{5-6} 
			& lowest \pt \phantom{xx}  &  highest \pt \phantom{xx}  & & lowest \pt \phantom{xx} &  highest \pt \phantom{xx} \\
			\midrule
			Yield extraction (\%) & 10 & 8 &  & 8 & 7 \\
			Tracking efficiency (\%) & 6 & 7 & & 3 & 5 \\
			Selection efficiency (\%) & 6 & 6 & & 3 & 3 \\
			PID efficiency (\%) & 5 & 5 & & 2 & 4 \\
			MC \pt shape (\%) &  negl. & negl.  & & negl. & negl. \\
			$(A\times\varepsilon)$ stat. unc. (\%)      & 1.7 & 1.8 & & 1.7 & 3.5 \\
			Beauty feed-down (\%)  & $\substack{+1.1 \\ -1.8}$ &  $\substack{+5.3 \\ -8.0}$ & & $\substack{+0.8 \\ -1.3}$ & $\substack{+2.6 \\ -4.0}$ \\       
			Branching ratio (\%) & \multicolumn{2}{c}{5.1} & & \multicolumn{2}{c}{5.0} \\
			Luminosity (\%) & \multicolumn{5}{c}{2.1} \\
			\midrule
		\end{tabular}
	\end{center}
	\caption{Summary of the systematic uncertainties for the two \Lc decay modes in \pp collisions at \sqrtsfive. The uncertainty sources found to be $< 1\%$ were considered negligible (``negl.'' in the table).}
	\label{tab:Syspp}

	\begin{center}
		\def\arraystretch{1.2}\tabcolsep=2pt    
		\begin{tabular}{lccccc} 
			\toprule
			& \multicolumn{2}{c}{$\LctopKpi$} & & \multicolumn{2}{c}{$\LctopKzS$} \\
			\cmidrule{2-3} \cmidrule{5-6} 
			& lowest \pt \phantom{xx} & highest \pt \phantom{xx} & & lowest \pt \phantom{xx} & highest \pt \phantom{xx}\\
			\midrule
			Yield extraction (\%) 	   & 8 & 10 &  & 10 & 8 \\
			Tracking efficiency (\%) & 6 & 6 & & 6 & 5 \\
			Selection efficiency (\%)         & 10 & 6 & & 15 & 8 \\
			PID efficiency (\%)          & 5 & 5 & & negl. & negl. \\
			MC \pt shape (\%)          & 1 & 1  & & 1 & 1 \\
			$(A\times\varepsilon)$ stat. unc. (\%)      & 1.1 & 4.0 & & 0.5 & 3.0 \\
			Beauty feed-down (\%)  & $\substack{+1.8 \\ -3.0}$ &  $\substack{+4.2 \\ -6.7}$ & & $\substack{+0.9 \\ -1.5}$ & $\substack{+4.6 \\ -7.0}$ \\       
			Branching ratio (\%) & \multicolumn{2}{c}{5.1} & & \multicolumn{2}{c}{5.0}\\
			Luminosity (\%) & \multicolumn{5}{c}{3.7} \\
			\midrule
		\end{tabular}
	\end{center}
	\caption{Summary of the systematic uncertainties for the two \Lc decay modes in \pPb collisions at \sqrtsNNfive. The uncertainty sources found to be $< 1\%$ were considered negligible (``negl.'' in the table).}
	\label{tab:SyspPb}
\end{table}  
\clearpage
\section{Results}
\label{sec: Results}

\subsection{\pt-differential cross sections}

\begin{figure}[ht!]
	\centering
	\includegraphics[width=0.48\textwidth]{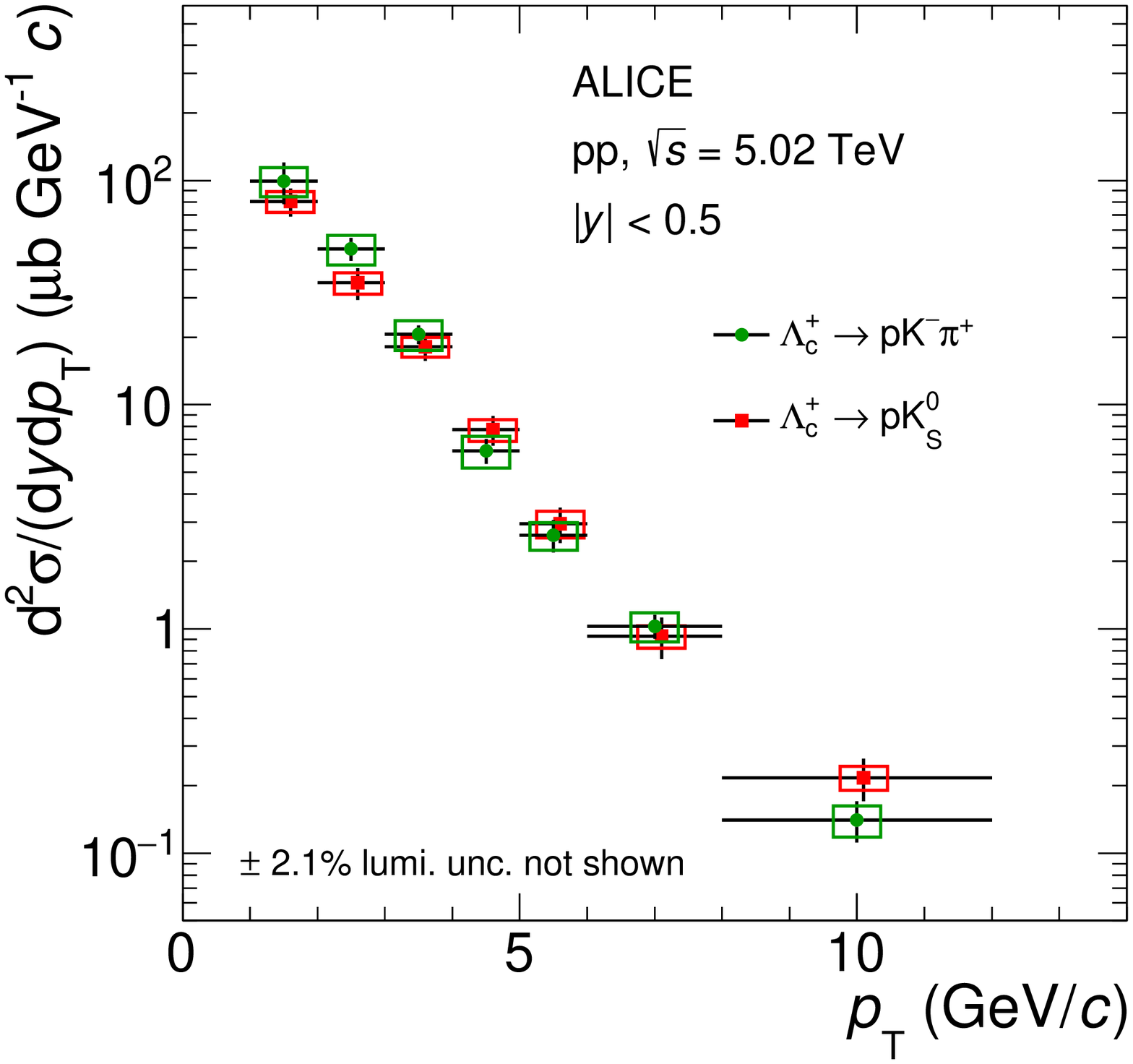}	\includegraphics[width=0.48\textwidth]{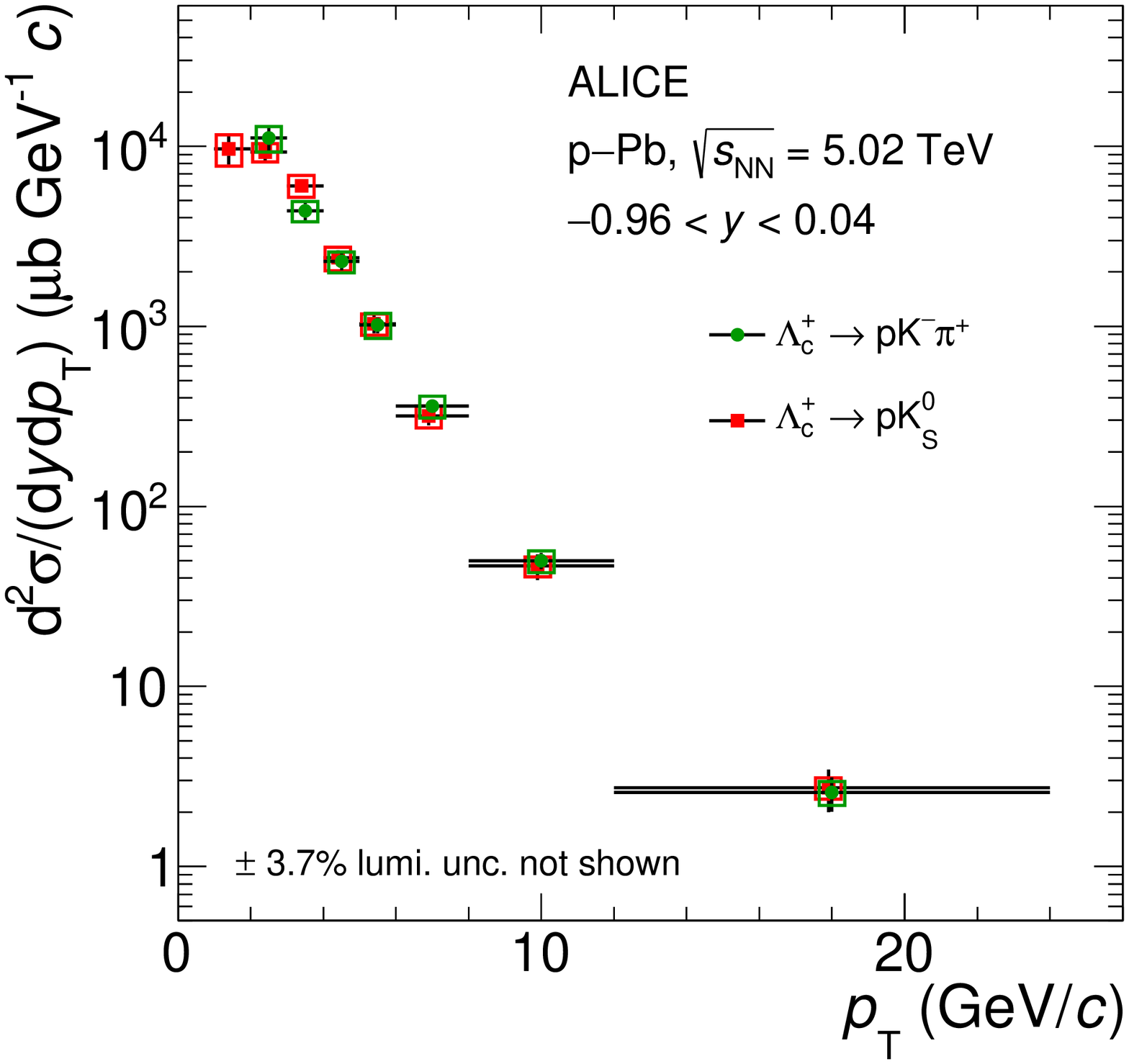}	
	\caption{Left: \pt-differential prompt \Lcplus-baryon cross section in \pp collisions at \sqrtsfive in the interval $1 < \pt < 12$ \gevc. Right: \pt-differential prompt \Lcplus-baryon cross section in \pPb collisions at \sqrtsNNfive in the interval $1 < \pt < 24$ \gevc. The statistical uncertainties are shown as vertical bars and the systematic uncertainties are shown as boxes. Horizontal position of points are shifted to provide better visibility.}
	\label{fig:CrossSection}
\end{figure}

The \pt-differential cross section of prompt \Lc-baryon production in pp collisions at \sqrtsfive, measured in the rapidity interval $\absrap < 0.5$ and $\pt$ interval $1 < \pt < 12$ \gevc, is shown in \figref{fig:CrossSection} (left)
 for the two decay channels \LctopKpi and \LctopKzS. \Figref{fig:CrossSection} (right) shows the \pt-differential cross section of prompt \Lc-baryon production in \pPb collisions at \sqrtsNNfive, measured in the rapidity interval $-0.96 < \rap < 0.04$ and \pt interval $1 <  \pt < 24$\,\gevc for the two decay channels \LctopKpi and \LctopKzS. The measurements in the different decay channels agree within statistical and uncorrelated systematic uncertainties, with the largest discrepancies among the measured values being smaller than $1.4\sigma$.

To obtain a more precise measurement of the \pt-differential \Lc-baryon production cross section, the results from the two decay channels were combined, taking into account the correlation between the statistical and systematic uncertainties. The systematic uncertainties treated as uncorrelated between the different decay channels (\LctopKpi and \LctopKzS) include those due to the raw-yield extraction, the \Lc-selection efficiency, and the $(A\times\epsilon)$ statistical uncertainties. The systematic uncertainties due to the
tracking efficiency, the PID efficiency, the generated \Lc \pt spectrum, the beauty feed-down, and the luminosity were treated as correlated between the two decay channels.
The branching ratio uncertainties were considered to be partially correlated, as described in~\cite{PDG20}.
A weighted average of the cross section values obtained from the different analyses was calculated, using the inverse of the quadratic sum of the relative statistical and uncorrelated systematic uncertainties as weights.

\begin{figure}[ht!]
	\centering
	\includegraphics[width=0.48\textwidth]{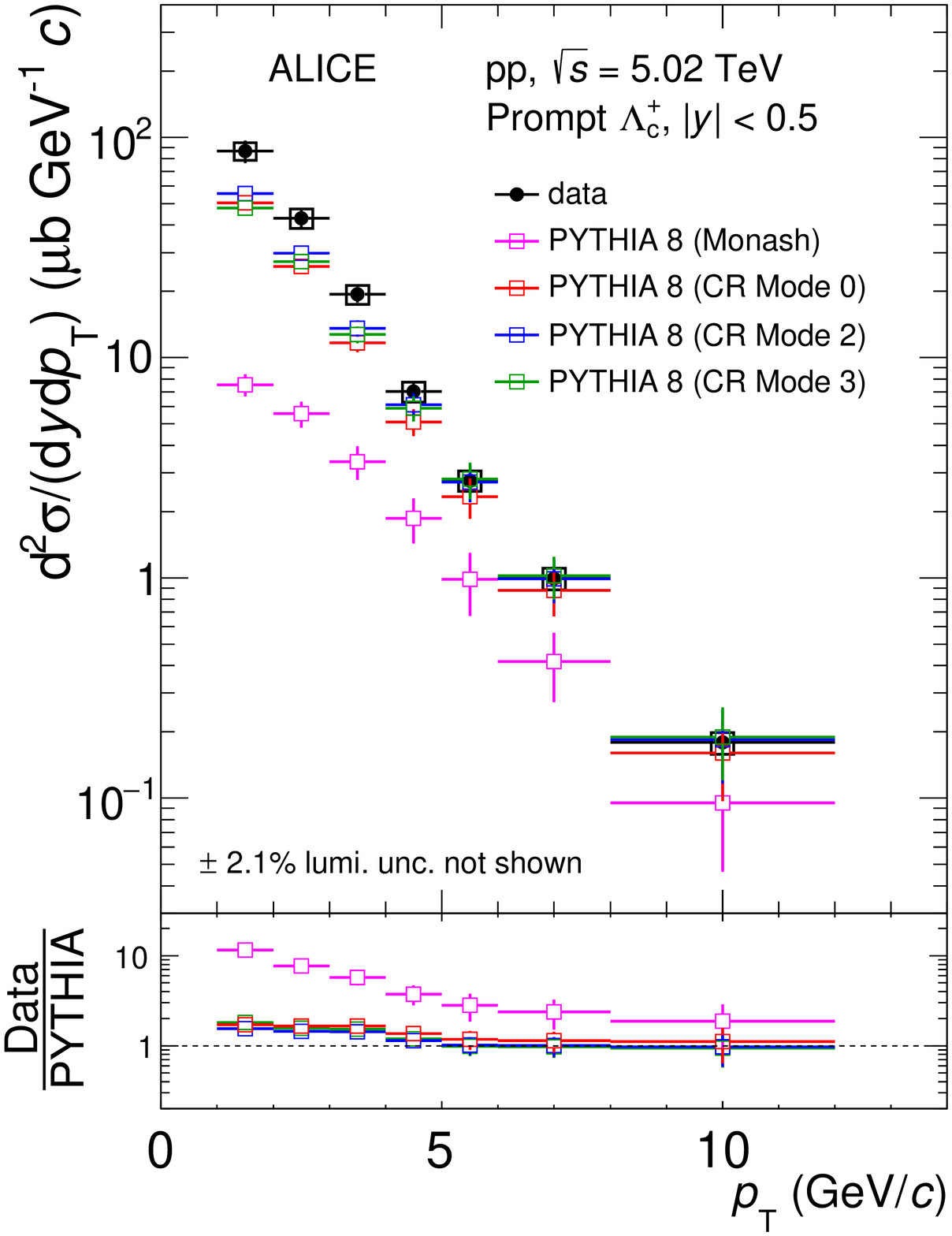}	
	\includegraphics[width=0.48\textwidth]{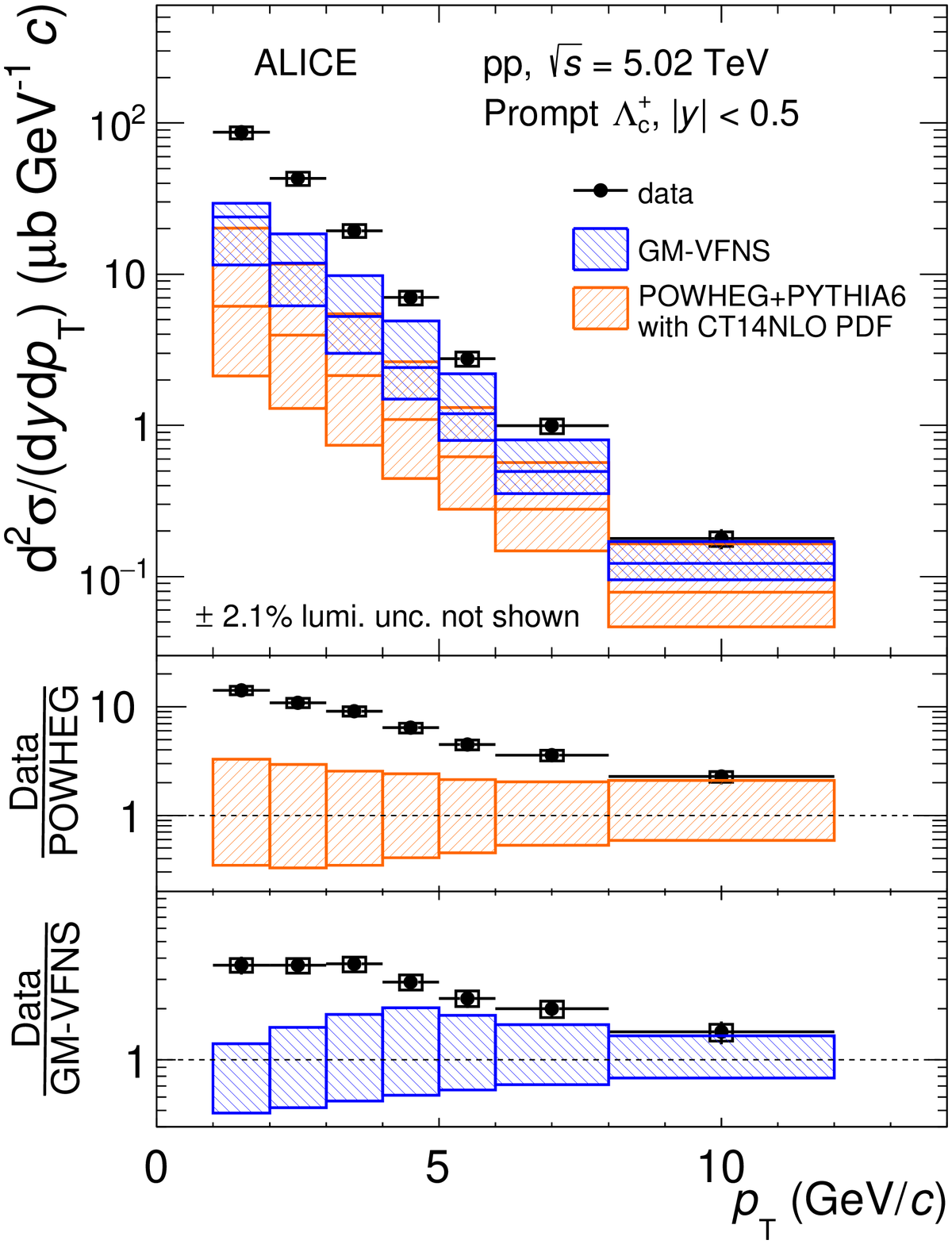}	
	\caption{Prompt \Lcplus-baryon \pt-differential production cross section in \pp collisions at \sqrtsfive in the interval $1 < \pt < 12~\gevc$. The statistical uncertainties are shown as vertical bars and the systematic uncertainties are shown as boxes. Left: Comparison to predictions from different tunes of the \mbox{PYTHIA 8} event generator~\cite{Skands:2014pea}~\cite{Christiansen:2015yqa}. The vertical bars on the \mbox{PYTHIA 8} predictions represent the statistical uncertainty from the simulation, and the vertical bars on the ratios in the bottom panel also include the statistical uncertainties from the data. Right: Comparison to predictions from the POWHEG event generator~\cite{Frixione:2007nw} and GM-VFNS calculations~\cite{Kniehl:2020szu}. The orange(blue) boxes represent the uncertainties of POWHEG(GM-VFNS) due to the choice of pQCD scales. See text for details on the \mbox{PYTHIA 8} and POWHEG event generator settings.}
	\label{fig:CrossSectionMergedMC}
\end{figure}

\Figref{fig:CrossSectionMergedMC} shows the measured production cross section (average of the two decay channels) in \pp collisions compared to predictions from MC generators and pQCD calculations. The left panel shows the comparison with predictions from different tunes of the \mbox{PYTHIA 8} generator, including the Monash tune~\cite{Skands:2014pea}, and tunes that implement colour reconnection (CR) beyond the leading-colour approximation~\cite{Christiansen:2015yqa}. These additional colour reconnection topologies include `junctions' which fragment into baryons, leading to increased baryon production. For the CR tunes, three modes are considered (Mode 0, 2, and 3), as described in~\cite{Christiansen:2015yqa}, which apply different constraints on the allowed reconnection, taking into account causal connection of dipoles involved in a reconnection and time-dilation effects caused by relative boosts between string pieces. It is noted that Mode 2 is recommended in~\cite{Christiansen:2015yqa} as the standard tune, and contains the strictest constraints on the allowed reconnection. In the simulations with the three CR modes, all soft QCD processes are switched on. All \mbox{PYTHIA 8} tunes underestimate the measured \pt-differential prompt \Lc cross section. The Monash tune significantly underestimates the cross section by a factor ${\sim}12$ for $1<\pt<2~\GeVc$, and around a factor 2--3 for $\pt>5~\GeVc$. All three CR modes yield a similar magnitude and shape of the \Lc cross section, and predict a  significantly larger \Lc production cross section with respect to the Monash tune. However, for all three CR modes, the measured \Lc production cross section is underestimated by a factor of about two for $1 < \pt < 2~\gevc$. For $\pt>5~\gevc$, Mode 2 and Mode 3 provide a good description of the data, while Mode 0 underestimates the data by 15--20\%. 
All tunes exhibit a harder \pt distribution than observed in data.

The right panel of \figref{fig:CrossSectionMergedMC} shows a comparison with a NLO pQCD calculation obtained with the POWHEG framework~\cite{Frixione:2007nw}, matched with \mbox{PYTHIA 6} to generate the parton shower, and the CT14NLO parton distribution functions~\cite{Dulat:2015mca}.
The nominal factorisation and renormalisation scales, $\mu_{\mathrm{F}}$ and $\mu_{\mathrm{R}}$, were taken to be equal to the transverse mass of the quark, $\mu_0 = \sqrt{m^2 + \pt^2}$, and the charm-quark mass was set to $m_\mathrm{c}=1.5~\gevcc$. The theoretical uncertainties were estimated by varying these scales in the range $0.5\mu_0 < \mu_{\mathrm{R,F}} < 2.0\mu_0$, with $0.5\mu_0 < \mu_{\mathrm{R}} / \mu_{\mathrm{F}} < 2.0\mu_0$. 
Results are also compared with recent GM-VFNS pQCD calculations~\cite{Kniehl:2020szu}. With respect to previous GM-VFNS calculations~\cite{Kniehl:2005mk,Kniehl:2012ti}, a new fragmentation function for \Lc has been used, obtained from a fit to OPAL data~\cite{Alexander:1996wy} and measurements from Belle at $\sqrt{s}$ = 10.52 \gev~\cite{Niiyama:2017wpp}.
The measured \pt-differential cross section is significantly underestimated by the POWHEG prediction, by a factor of up to 15 in the lowest \pt interval of the measurements, and around a factor 2.5 in the highest. While the discrepancy between the data and calculation decreases as the $\pt$ increases, the measured cross section at $8 < \pt < 12~\gevc$ is still ${\sim}50\%$ larger than the upper edge of the POWHEG uncertainty band. The discrepancy between the data and POWHEG is similar to what was observed in pp collisions at $\sqrts = 7~\tev$~\cite{Acharya:2017kfy}. The GM-VFNS predictions also significantly underestimate the data, by about a factor of 3--4 at low \pt and by about a factor of 1.5 at high \pt. 
 	
\begin{figure}[ht!]
		\centering
		\includegraphics[width=0.48\textwidth]{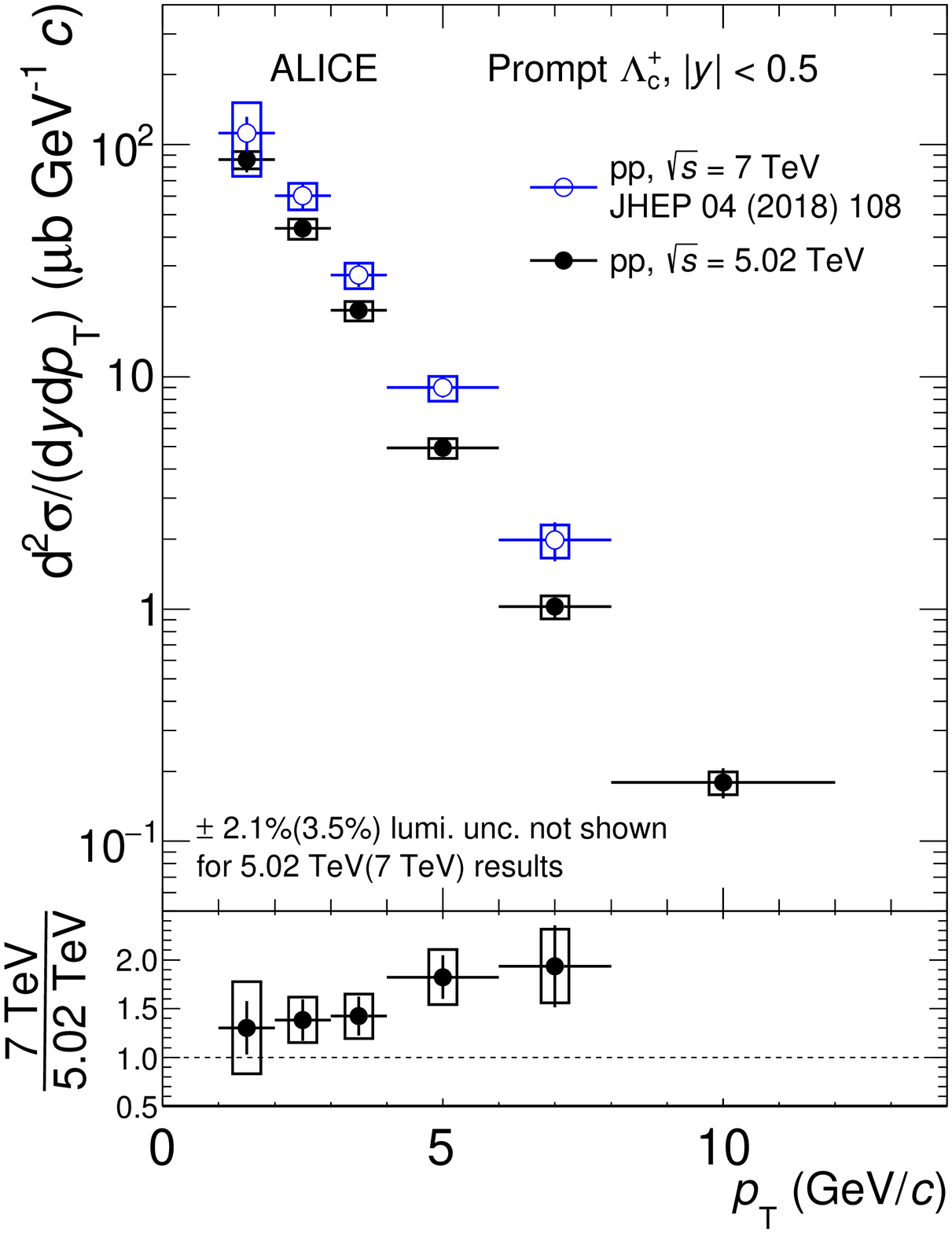}
		\caption{Comparison between the \pt-differential production cross section of prompt \Lcplus baryons in \pp collisions at $\sqrt{s}=7~\tev$~\cite{Acharya:2017kfy} and \sqrtsfive. 
		The ratio between the cross sections is shown in the lower panel.
The statistical uncertainties are shown as vertical bars and the systematic uncertainties are shown as boxes.}
		\label{fig:5vs7CrossRatio}
	\end{figure}

In \figref{fig:5vs7CrossRatio}, the \Lc-production cross section in \pp collisions at \sqrtsfive is compared with the measurement at \sqrtsseven~\cite{Acharya:2017kfy}. For a direct comparison, the intervals $4 < \pt < 5$ \gevc and $5 < \pt < 6$ \gevc of the \sqrtsfive analysis have been merged. When merging, the systematic uncertainties were propagated considering the uncertainty due to the raw-yield extraction as fully uncorrelated and all the other sources as fully correlated between $\pt$ intervals. 
In the lower panel of the same figure, the ratio of the cross sections is shown. In this case, the systematic uncertainties on feed-down, \pt shape, and branching ratio were assumed to be fully correlated, while all the other sources were considered as uncorrelated between the results at the two collision energies.
The relative statistical uncertainties in the measurement at $\sqrts = 5.02~\tev$ are on average smaller than those in the measurement at $\sqrts = 7~\tev$ by a factor ${\sim}1.5$.
As expected, a lower \Lc-production cross section is observed at the lower collision energy. The difference between the cross sections at the two \sqrts values increases with increasing \pt, indicating a harder \pt shape at the higher collision energy. This behaviour is consistent with that observed for the D-meson cross section ratios at $\sqrt{s}=7~\tev$ and \sqrtsfive, which is described by pQCD calculations~\cite{Acharya:2019mgn}.

	\begin{figure}[ht!]
	\centering
	\includegraphics[width=0.48\textwidth]{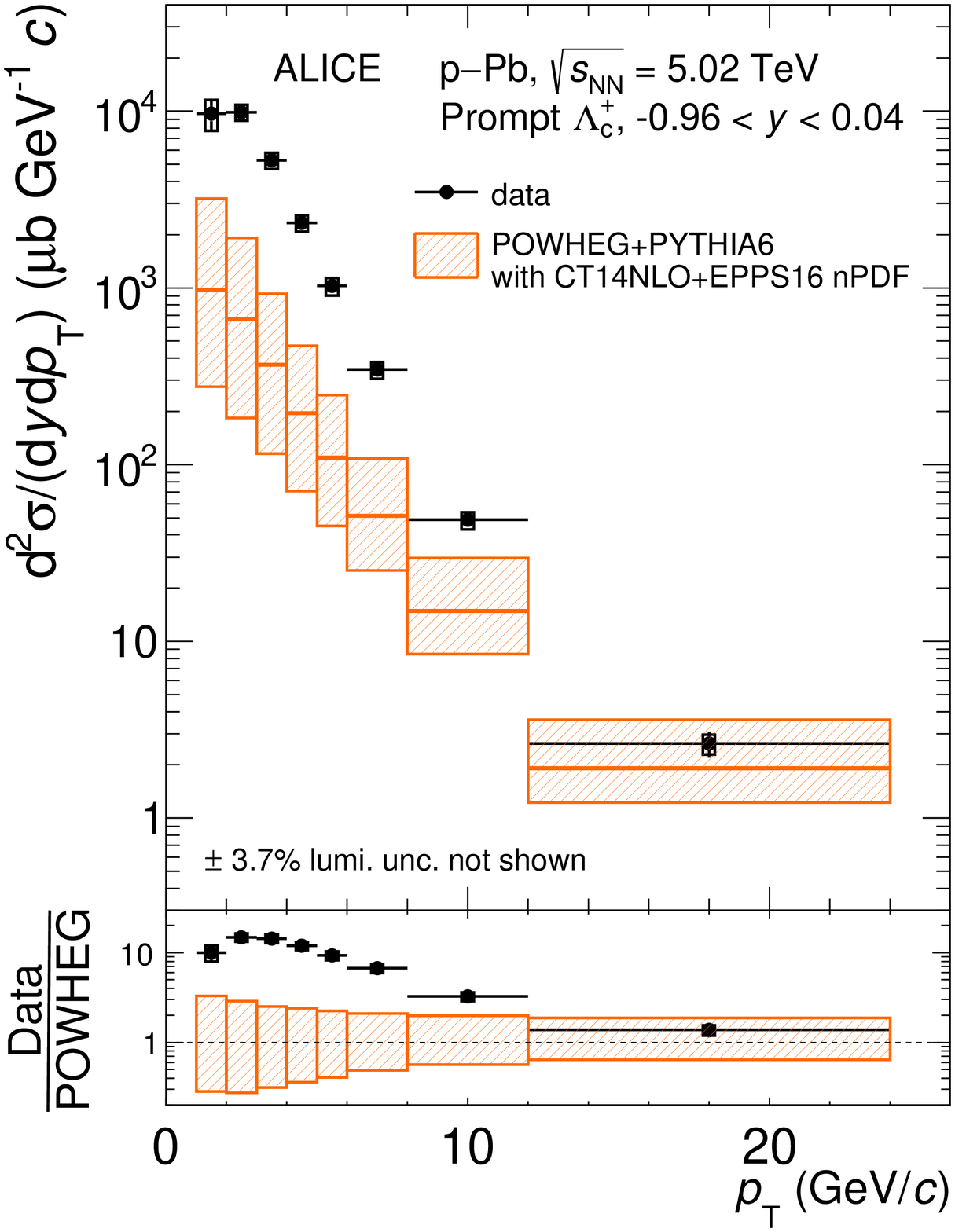}	
	\caption{\pt-differential prompt \Lcplus-baryon production cross section in \pPb collisions at \sqrtsNNfive in the interval $1 < \pt < 24~\gevc$ compared to predictions from the POWHEG event generator~\cite{Frixione:2007nw}. The statistical uncertainties are shown as vertical bars and the systematic uncertainties are shown as boxes. The orange boxes represent the uncertainties of POWHEG due to the choice of pQCD scales. See text for details on the POWHEG event generator settings. }
	\label{fig:CrossSectionMerged}
\end{figure}

\Figref{fig:CrossSectionMerged} shows the $\pt$-differential cross section averaged among the decay channels and analysis techniques in \pPb collisions. The cross section is compared to the POWHEG event generator, where the generator settings, the parton shower, and the set of parton distribution functions are the same as used in the calculations for \pp collisions, 
and the nuclear modification of the parton distribution functions is modelled with the EPPS16 nPDF parameterisation~\cite{Eskola:2016oht}.
The theoretical uncertainty includes the uncertainty on the factorisation and renormalisation scales (estimated as done for POWHEG predictions for \pp collisions), while the uncertainties on the parton distribution functions and EPPS16 nPDF are not included in the calculation as they are smaller than the scale uncertainties. 
The cross section is underestimated by the POWHEG prediction by a factor of up to 15 in the lowest \pt intervals, similar to what is observed for \pp collisions. The difference between the POWHEG predictions and the measured cross section decreases with increasing \pt and in the highest \pt interval of the
measurement ($12 < \pt < 24~\gevc$) the data point lies on the upper edge of the POWHEG uncertainty band.
The Run 2 \pPb results are compatible with our previous results from the sample of \pPb collisions at \sqrtsNNfive collected in LHC Run 1~\cite{Acharya:2017kfy}.
The statistical uncertainties have been reduced by approximately a factor of two for all \pt intervals, and the systematic uncertainties improved by approximately 30\% at low \pt and 10\% at high \pt. 

\subsection{Nuclear modification factor}

\begin{figure}[ht!]
	\centering
	\includegraphics[width=0.95\textwidth]{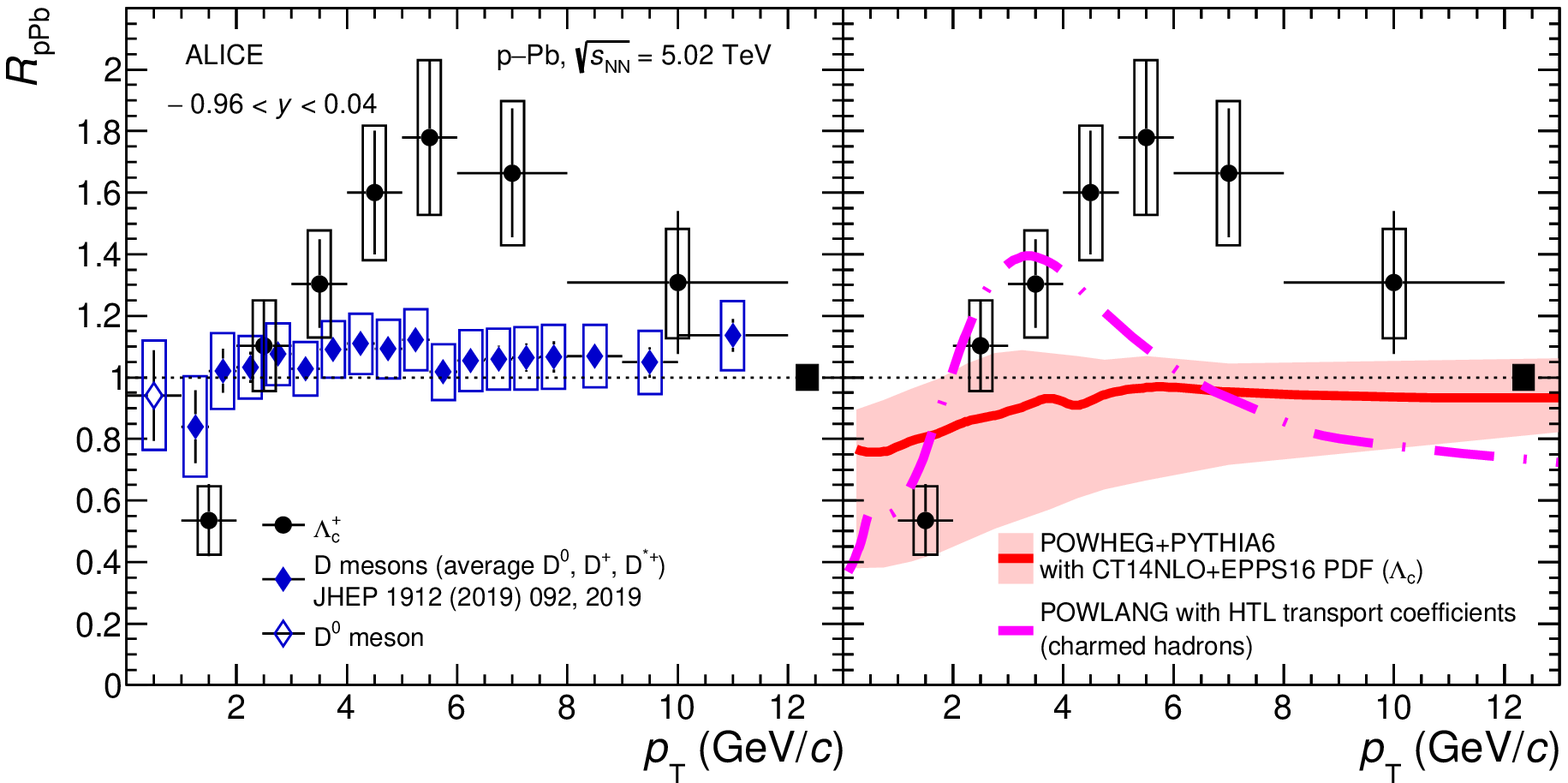}
	\caption{The nuclear modification factor \RpPb of prompt \Lcplus baryons in \pPb collisions at \sqrtsNNfive as a function of \pt, compared to the \RpPb of D mesons~\cite{Acharya:2019mno} (average of \DZero, \Dplus, and \Dstar in the range $1 < \pt < 12$ \gevc and \DZero in $ 0 < \pt < 1$ \gevc) (left), as well as to POWHEG+PYTHIA 6~\cite{Frixione:2007nw} with EPPS16~\cite{Eskola:2016oht} simulations, and POWLANG~\cite{Beraudo:2015wsd} predictions (right). The black-filled box at $\RpPb = 1$ represents the normalisation uncertainty.}
	\label{fig:RpPb}
\end{figure}

The nuclear modification factor \RpPb was calculated as the \pt-differential $\Lc$ cross section in \pPb collisions divided by the reference measurement of the \pt-differential $\Lc$ cross section in \pp collisions scaled by the lead mass number $A = 208$

\begin{equation}	
R_{\mathrm{pPb}} = \frac{1}{A} \frac{\mathrm{d}\sigma_{\mathrm{pPb}}/\mathrm{d}\pt}
{ \mathrm{d}\sigma_{\mathrm{pp}}/\mathrm{d}\pt} 
\label{eq:RpPbScaled}
\end{equation}

where $\mathrm{d}\sigma_\mathrm{pp}/\mathrm{d}\pt$ was obtained from the cross section measured in \pp collisions in $|y|<0.5$ applying a correction factor to account for the different rapidity coverage of the \pp and \pPb measurements. 
The correction factor is calculated with FONLL and ranges from 0.995 (in $1 < \pt < 2~\gevc$) to 0.983 (in $8 < \pt < 12~\gevc$).
\Figref{fig:RpPb} (left) shows the \RpPb of \Lc baryons in the \pt interval $1 < \pt < 12$ \gevc compared to the \RpPb of non-strange D mesons from~\cite{Acharya:2019mno}. With respect to the previous measurement of the \Lc-baryon \RpPb~\cite{Acharya:2017kfy}, the \pt reach has been extended to higher and lower \pt. In addition, the \pp reference at the same per-nucleon centre-of-mass energy as the \pPb sample eliminates the uncertainty originating from the $\sqrt{s}$-scaling of the \pp cross section measured at \sqrtsseven that was present in the previous results. These improvements, along with the increased statistical precision, have allowed for a reduction of the overall uncertainty of the \RpPb by a factor of 1.7--2 compared with the previous measurement.  The result is consistent with the D-meson \RpPb 
within the uncertainties in the \pt regions $1 < \pt< 4$ \gevc and $\pt>8$ \gevc, but larger than the D-meson \RpPb in $4 < \pt < 8$ \gevc with a maximum deviation of 1.9$\sigma$ in $5 < \pt < 6~\gevc$, where $\sigma$ is defined as the quadratic sum of the statistical and the lower(upper) systematic uncertainties for $\Lc$ baryons (D mesons). 
For $\pt > 2~\gevc$ the \Lc-baryon \RpPb is systematically above unity, with a maximum deviation from $\RpPb=1$ reaching 2.2$\sigma$ in the \pt interval $5 < \pt < 6$ \gevc, where $\sigma$ is defined as the quadratic sum of the statistical and the upper systematic uncertainty.
In the \pt interval $1 < \pt< 2$ \gevc the \RpPb is lower than unity by 2.6$\sigma$. This hints that \Lc production is suppressed at low \pt and is enhanced at mid-\pt in \pPb collisions with respect to pp collisions. 
In \figref{fig:RpPb} (right)  the measured \Lc-baryon \RpPb is compared to model calculations. The POWHEG+PYTHIA 6 simulations use the POWHEG event generator with PYTHIA 6 
parton shower and EPPS16 parameterisation of the nuclear modification of the PDFs~\cite{Eskola:2016oht}. The uncertainty band includes the uncertainties on the nuclear PDFs and on the choice of the pQCD scales. 
The POWLANG model~\cite{Beraudo:2015wsd} assumes that a hot deconfined medium is formed in \pPb collisions, and the transport of heavy quarks through an expanding QGP is computed utilising the Langevin approach and Hard Thermal Loop (HTL) transport coefficients. The POWLANG model does not implement specific differences in hadronisation mechanisms for baryons and mesons, and the same prediction holds for all charm hadron species. The two models capture some features of the data, but neither of them can quantitatively reproduce the observed \Lc-baryon \RpPb in the measured \pt interval.

\subsection{\pt-integrated \Lc cross sections}

The visible \Lc cross section  
was computed by integrating the \pt-differential cross section in its measured range.
In the integration, the systematic uncertainties were propagated considering the uncertainty due to the raw-yield extraction as fully uncorrelated and all the other sources as fully correlated between $\pt$ intervals.
The visible \Lc cross section in \pp collisions at \sqrtsfive is

\begin{equation}
\mathrm{d}\sigma^{\Lc}_{{\rm pp,~5.02~TeV}}/{\mathrm{d}y|^{1 < \pt < 12~\gevc}_{|y|<0.5}} =  161 \pm 11 {\rm~(stat.)} \pm 14 {\rm~(syst.)} \pm 3 {\rm~(lumi.)}~\mub .
\label{eq:pp-vis}
\end{equation}

The visible \Lc cross section in \pPb collisions is

\begin{equation}
\mathrm{d}\sigma^{\Lc}_{{\rm pPb,~5.02~TeV}}/{\mathrm{d}y|^{1 < \pt < 24~\gevc}_{-0.96<y<0.04}}  =  29.0 \pm 2.0 {\rm~(stat.)} \pm 3.6 {\rm~(syst.)} \pm 1.1 {\rm~(lumi.)}~\mathrm{mb} .
\label{eq:pPb-vis}
\end{equation}

The \pt-integrated \Lc production cross section at midrapidity was obtained by extrapolating the visible cross sections to the full \pT range. 
The extrapolation approach used for D mesons~\cite{Acharya:2017jgo}, based on the \pt-differential cross sections predicted by FONLL calculations, is not applicable here because FONLL does not have predictions for \Lc baryons.
For \pp collisions, \mbox{PYTHIA 8} predictions with specific tunes implementing CR mechanisms were used for the extrapolation. The \pt-differential \Lc cross section values in $0 < \pt < 1$ \gevc and for \pt $\ge$ 12 \gevc were obtained by scaling the measured \Lc cross section in $1 < \pt < 12$ \gevc for the fractions of cross section given by PYTHIA in $0 < \pt < 1$ \gevc and for \pt $\ge$ 12 \gevc respectively. 
The \mbox{PYTHIA 8} simulation with Mode 2 CR tune~\cite{Christiansen:2015yqa} including soft QCD processes, which gives the best description of both the magnitude and shape of the \Lc cross section and \LcD ratio, was used to calculate the central value of the extrapolation factors. 
The procedure was repeated considering the three modes defined in~\cite{Christiansen:2015yqa}, with the envelopes of the corresponding results assigned as the extrapolation uncertainty.
A second extrapolation method was also implemented as a cross check. This consisted of multiplying the measured \DZero cross section value in $0 < \pT < 1~\gevc$ by the \LcD ratio estimated with \mbox{PYTHIA 8} (CR Mode 2) in the same \pt interval to get an estimate of the \Lc cross section value in $0 < \pT < 1~\gevc$, and then integrating in \pt. The results obtained with the two methods were found to be compatible within the uncertainties. 

The resulting \pt-integrated cross section of the \Lc baryon in pp collisions at \sqrtsfive is
\begin{equation}
\begin{split}
{\mathrm{d}\sigma^{\Lc}_{\rm pp,~5.02~TeV}}/{\mathrm{d}y|_{|y|<0.5}} =~230 \pm 16 {\rm~(stat.)~} \pm 20 {\rm~(syst.)~} \pm 5 {\rm~(lumi.) \substack{+5\\ -10}~} {\rm~(extrap.)} \,\mub&.
\end{split}
\end{equation}

In \pPb collisions, the \pt-integrated \Lc-production cross section was obtained using a different approach, since the \pt spectrum of \Lc is not well described by PYTHIA or other event generators. In this case, the cross sections in $0 < \pt < 1~\gevc$ and $\pt>24~\gevc$ were calculated as the product of the \pp cross sections in these $\pt$ intervals obtained from the extrapolation of the measured \pt-differential cross section, as described above; the Pb mass number; a correction factor to account for the different rapidity interval covered in \pp and \pPb collisions; and an assumption on the nuclear modification factor \RpPb as described hereafter. 
For $0 < \pt < 1~\gevc$, the \RpPb was taken as $\RpPb = 0.5$ as in the $ 1 < \pT < 2~\gevc$ interval, under the hypothesis that the trend of the \Lc \RpPb at low \pt is similar to that of D mesons.
The uncertainty was estimated by varying the hypothesis in the range $0.35 < \RpPb < 0.8$, which incorporates the envelope of the available models (see \figref{fig:RpPb}) and the range defined by the combination of the statistical and systematic uncertainties of the \Lc \RpPb in $1 < \pt < 2~\gevc$.
For $ \pT > 24$ \gevc, the \RpPb was assumed to be equal to unity, with the range $0.8 < \RpPb < 1.2$ used to define the uncertainty.

The resulting \pt-integrated cross section of prompt \Lc in \pPb collisions at \sqrtsNNfive is
\begin{equation}
\begin{split}
{\mathrm{d}\sigma^{\Lc}_{\rm pPb,~5.02~TeV}}/{\mathrm{d}y|_{-0.96<y<0.04}} =~36.2 \pm 2.5  {\rm~(stat.)~} \pm 4.5 {\rm~(syst.)~}  \pm 1.3 {\rm~(lumi.) \substack{+4.4\\ -2.7}~} {\rm~(extrap.)} \,\rm mb&.
\end{split}
\end{equation}
The visible cross sections make up $70\%$ and $80\%$ of the integrated cross sections in pp and \pPb collisions, respectively. The \pt-integrated \Lc cross sections in \pp and \pPb collisions can be used for the comparison of fragmentation fractions of charm quarks in different collision systems and rapidity intervals. They 
can also be used in the calculation of the $\mathrm{c\bar{c}}$ cross section together with the cross sections of D mesons and higher-mass charm baryons that do not decay into \Lc. Due to the lack of measurements of higher-mass charm baryons ($\Xi_{\mathrm{c}}^{+,0}, \Omega_{\mathrm{c}}$) at \sqrtsfive, which contribute to the $\mathrm{c\bar{c}}$ cross section, a calculation of the $\mathrm{c\bar{c}}$ cross section is beyond the scope of this work.

\subsection{\LcD ratios}

The ratios between the yields of \Lc baryons and $\Dzero$ mesons were calculated using the \Dzero cross sections reported in~\cite{Acharya:2019mgn} for \pp collisions and~\cite{Acharya:2019mno} for \pPb collisions, respectively. The uncertainty sources assumed to be uncorrelated between the \Lc and \Dzero production cross sections include those due to the raw-yield extraction, the selection efficiency, the PID efficiency, the generated \pt shape, the $(A\times\epsilon)$ statistical uncertainties, and the branching ratios. The uncertainties assumed to be correlated include those due to the tracking, the beauty feed-down and the luminosity. The $\Dzero$ cross section was measured in finer \pt intervals than the \Lc, so it was rebinned such that the \pt intervals match between the two species.

\begin{figure}[ht!]
	\centering
	\includegraphics[width=0.6\textwidth]{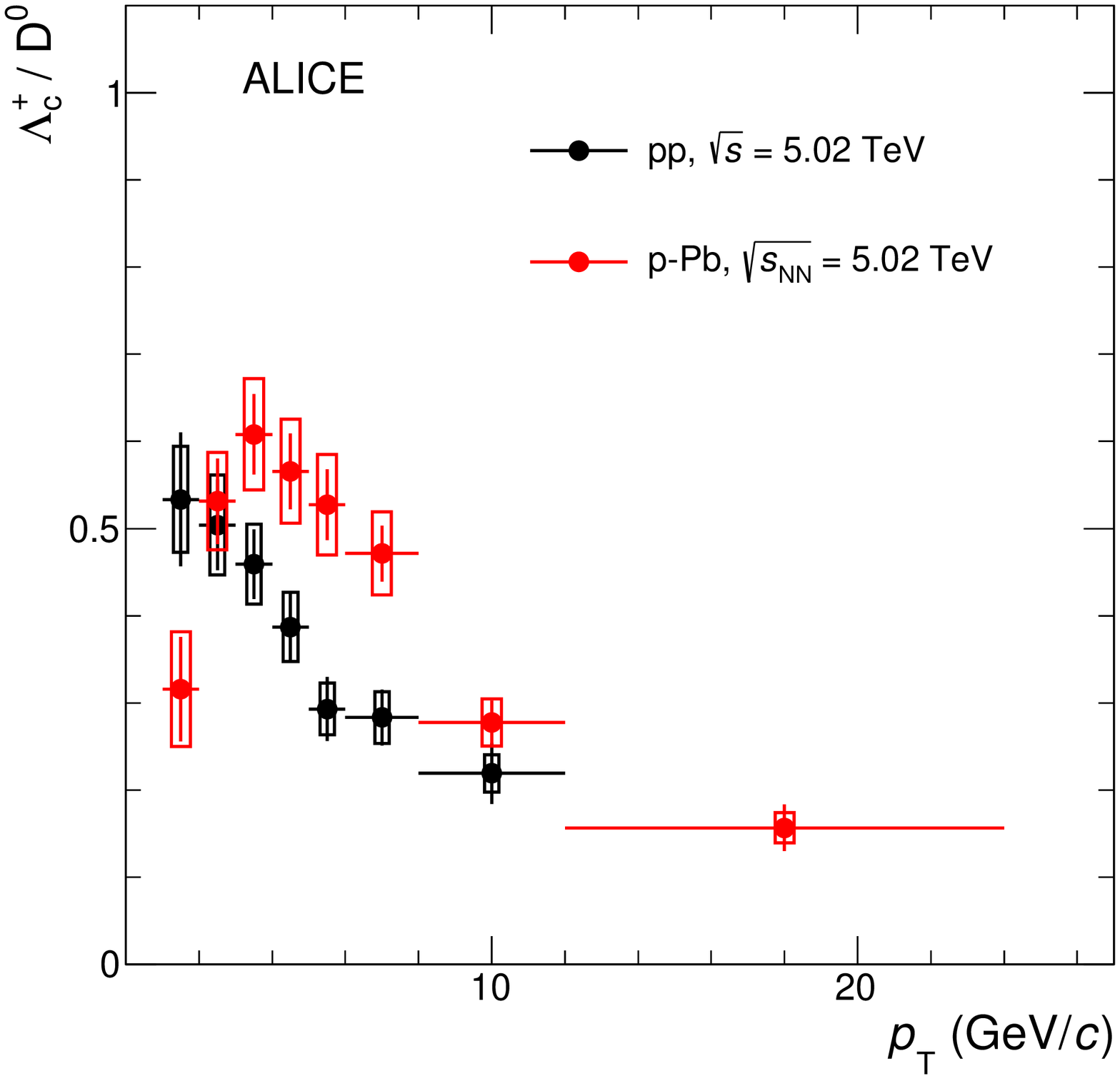}
	\caption{The \LcD ratio as a function of \pt measured in pp collisions at \sqrtsfive and in \pPb collisions at \sqrtsNNfive.}
	\label{fig:LcD-pp-pPb}
\end{figure}

The \LcD ratio as a function of \pt in \pp and \pPb collisions is shown in \figref{fig:LcD-pp-pPb}.
A clear decreasing trend with increasing $\pT$ is seen in both pp and \pPb collisions for $\pt > 2$ \gevc, and at high \pt the ratio reaches a value of about 0.2. 
The ratios measured in \pp and \pPb collisions are qualitatively consistent with each other, although 
a larger \LcD ratio in $3 < \pt < 8~\gevc$ and a lower ratio in $1 < \pt < 2~\gevc$ are measured in \pPb collisions with respect to \pp collisions.

The values of the \pt-integrated \LcD ratios are reported in~\tabref{tab:lcd0ratio} along with the values measured in $\ee$ and $\ep$ collisions by other experiments. 
The \LcD ratios in \pp and \pPb collisions are consistent with each other within the experimental uncertainties.
Comparing to previous measurements in other collision systems, the \LcD ratio is significantly enhanced by a factor of about 3--5 in \pp collisions and a factor of about 2--4 in \pPb collisions, indicating that the fragmentation fractions of charm quarks into baryons are different with respect to $\ee$ and $\ep$ collisions.
This is consistent with the previous ALICE measurements~\cite{Acharya:2017kfy}, where the \pt-integrated \LcD ratios were restricted to $1 < \pt < 8~\gevc$ in \pp collisions, and to $2 < \pt < 12~\gevc$ in \pPb collisions.

\begin{table}[h!]
	\small
	\begin{center}
		\def\arraystretch{1.4}\tabcolsep=6pt
		\begin{tabular}{lrccr}
			\toprule
			& \Lcplus/\Dzero $\pm$ stat. $\pm$ syst. & System & \sqrts~(GeV)& Notes \\
			\midrule
			ALICE & $0.51 \pm 0.04 \pm 0.04~\substack{+0.01\\ -0.02}$ & pp & 5020   & $\pt > 0, |y| < 0.5$ \\
			ALICE & $0.42 \pm 0.03 \pm 0.06 \substack{+0.05\\ -0.03}$ & \pPb & 5020   & $\pt > 0, -0.96 < y < 0.04$ \\
			CLEO~\cite{Avery:1990bc} & $0.119 \pm 0.021 \pm 0.019$ & \ee & $10.55$   & \\
			ARGUS~\cite{Albrecht:1988an,Albrecht:1991ss} & $0.127 \pm 0.031$ & \ee & $10.55$  & \\
			LEP average~\cite{Gladilin:2014tba} & $0.113 \pm 0.013 \pm 0.006$ & \ee & $91.2$  & \\
			\multirow{2}{*}{ZEUS DIS~\cite{Abramowicz:2010aa}} & \multirow{2}{*}{$0.124 \pm 0.034 \substack{ +0.025\\ -0.022}$} & \multirow{2}{*}{\ep} & \multirow{2}{*}{$320$}   & $1 < Q^2 < 1000$ $\gev ^2$, \\
			& & & & $0 < \pt < 10$ \gevc, $0.02 < y < 0.7$\\[0.2cm]
			ZEUS $\gamma$p,  & \multirow{2}{*}{$0.220 \pm 0.035 \substack{+0.027\\ -0.037} $} & \multirow{2}{*}{\ep} & \multirow{2}{*}{$320$} & $130 < W < 300$ \gev, $Q^2 < 1$ $\gev^2$, \\
			HERA I~\cite{Chekanov:2005mm} & & & & $\pt > 3.8$ \gevc, $|\eta| < 1.6$\\[0.2cm]
			ZEUS $\gamma$p,  & \multirow{2}{*}{$0.107 \pm 0.018 \substack{+0.009\\ -0.014} $} & \multirow{2}{*}{\ep} & \multirow{2}{*}{$320$} & $130 < W < 300$ \gev, $Q^2 < 1$ $\gev^2$,\\
			HERA II~\cite{Abramowicz:2013eja} & & & & $\pt > 3.8$ \gevc, $|\eta| < 1.6$\\[0.2cm]
			\midrule
		\end{tabular}
	\end{center}
	\caption{Comparison of the \pt-integrated \Lcplus/\Dzero ratio measured in \pp and \pPb collisions, and the same ratios in \ee and \ep collisions (reproduced from~\cite{Acharya:2017kfy}). Statistical and systematic uncertainties are reported (from references~\cite{Albrecht:1988an,Albrecht:1991ss} it was not possible to separate systematics and statistical uncertainties). The ALICE measurements report an additional uncertainty source from the extrapolation procedure. }
	\label{tab:lcd0ratio}
\end{table}

\begin{figure}[t!b]
	\centering
	\includegraphics[width=0.48\textwidth]{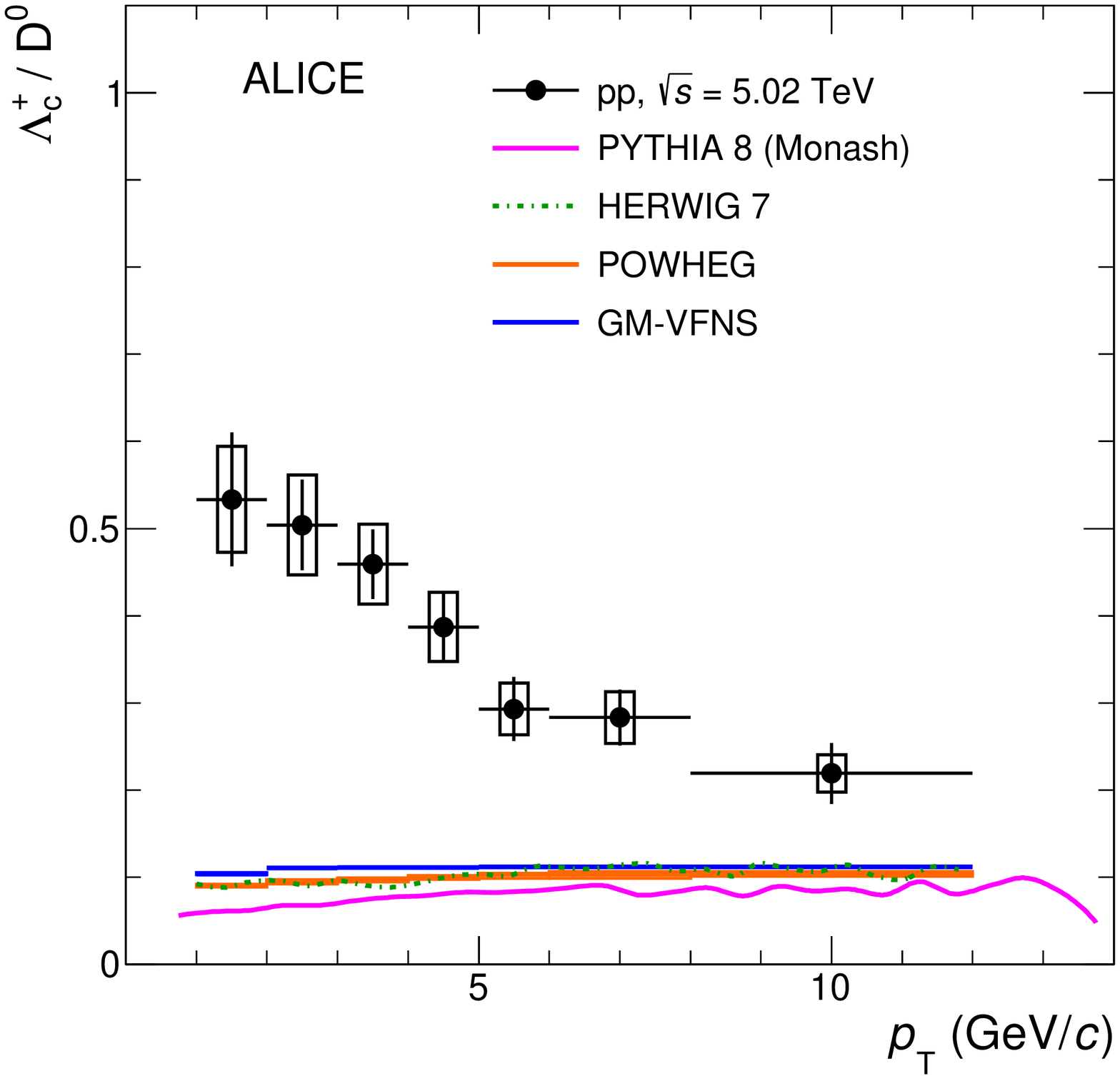}
	\includegraphics[width=0.48\textwidth]{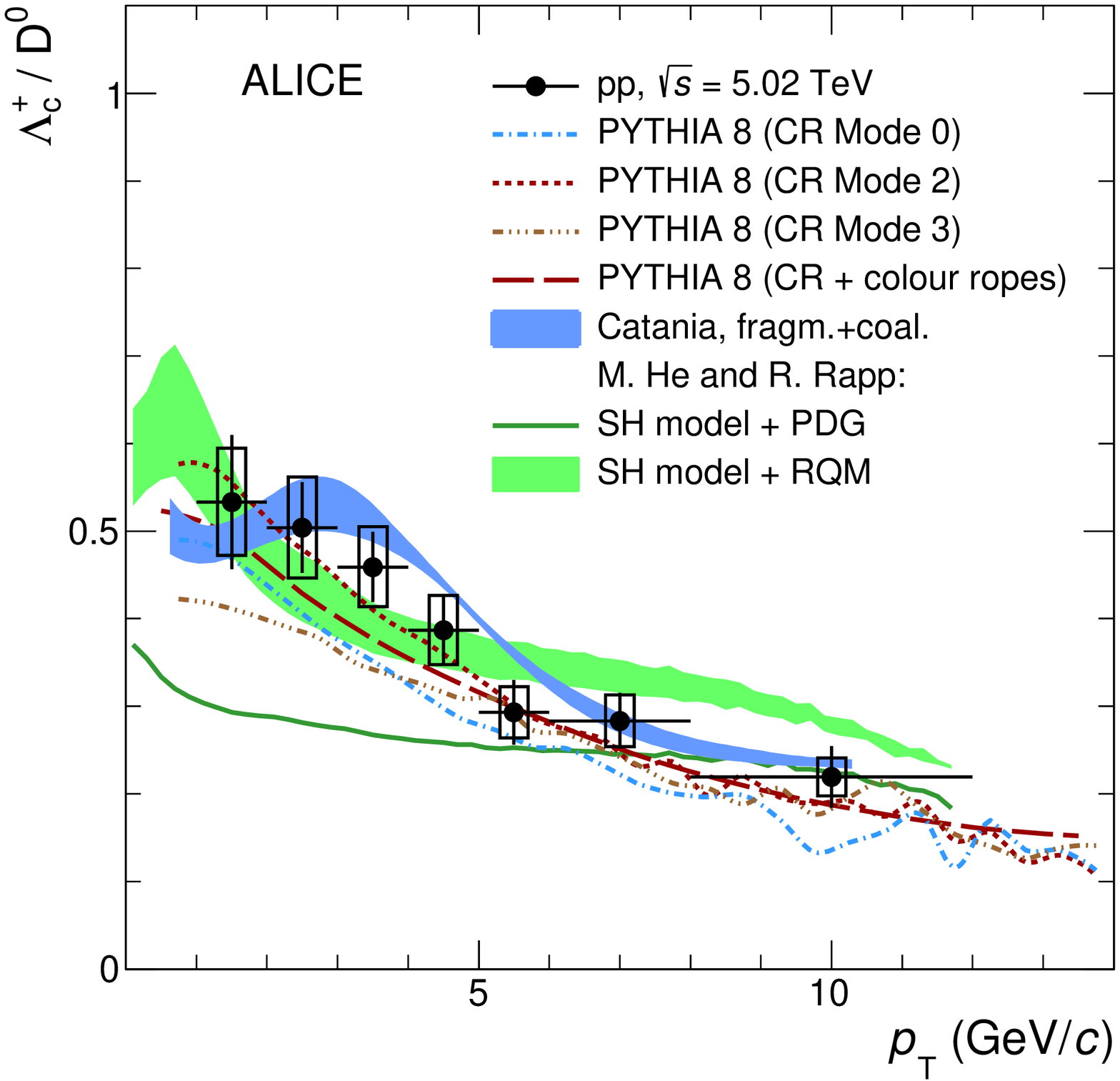}
	\caption{The \LcD ratio measured in pp collisions at \sqrtsfive, compared to theoretical predictions. The measurement is compared with predictions from MC generators (PYTHIA 8~\cite{Skands:2014pea,Christiansen:2015yqa}, HERWIG 7~\cite{Bellm:2015jjp}, 
	POWHEG~\cite{Frixione:2007nw}), GM-VFNS~\cite{Kniehl:2020szu}, a statistical hadronisation model~\cite{He:2019tik} (`SH model' in the legend) and a model which implements hadronisation via coalescence and fragmentation~\cite{Minissale:2020bif}. See text for model details.}
	\label{fig:LcD-models}
\end{figure}

\Figref{fig:LcD-models} shows the \LcD ratio in \pp collisions compared with models from MC generators, and a statistical hadronisation model. The MC generators include PYTHIA 8 with Monash tune and colour reconnection tunes as described above;
PYTHIA 8 with colour reconnection plus rope hadronisation~\cite{Bierlich:2014xba,Christiansen:2015yqa} where colour charges can act coherently to form a rope, increasing the effective string tension;
HERWIG 7.2~\cite{Bellm:2015jjp} where hadronisation is implemented via clusters; 
and POWHEG pQCD generator matched to \mbox{PYTHIA 6} to generate the parton shower, as described above. The measured points are also compared to predictions from GM-VFNS pQCD calculations, which were computed as the ratios of the \Lc and \Dzero cross sections obtained with the same choice of pQCD scales~\cite{Kniehl:2020szu}. 
The left panel shows the predictions of the \LcD ratio from PYTHIA 8 (Monash tune), HERWIG 7, POWHEG, and GM-VFNS, which all implement fragmentation processes tuned on charm production measurements in $\ee$ collisions, and therefore all predict a value of the \LcD ratio around 0.1, with a very mild \pt dependence. These predictions significantly underestimate the data at low \pt by a factor of about 5--10, while at high \pt the discrepancy is reduced to a factor of about 2. 
The right panel shows models which include processes that enhance baryon production.
A significant enhancement of the \LcD ratio is observed with PYTHIA 8 simulations including CR beyond the leading-colour approximation, with respect to the Monash tune. The results of these \mbox{PYTHIA 8} tunes are consistent with the measured \LcD ratio in pp collisions, also reproducing the decreasing trend of \LcD with increasing \pt. 
Including rope hadronisation in addition to
colour reconnection induces a small modification in the \LcD ratio, suggesting that the increased string tension does not significantly affect the relative production of baryons with respect to mesons.
The data is also compared with a statistical hadronisation model~\cite{He:2019tik} where the underlying charm baryon spectrum is either taken from the PDG, or augmented to include additional excited baryon states, which have not yet been observed but are predicted by the Relativistic Quark Model (RQM)~\cite{Ebert:2011kk}. For the former case, the model underpredicts the data at low \pt. For the latter case, the additional charm baryon states decay strongly to \Lc baryons, contributing to the prompt \Lc spectrum. This increases the \LcD ratio and allows the model to describe both the magnitude and the \pt dependence of the measured ratio.
Finally, the Catania model~\cite{Minissale:2020bif} is also presented, which assumes that a QGP is formed in \pp collisions and that the hadronisation occurs via coalescence as well as fragmentation. The light quark \pt spectrum is determined with a blast wave model, while the heavy quark \pt spectrum is determined with FONLL pQCD predictions, and coalescence is implemented via the Wigner formalism. Contrary to the implementation in \PbPb collisions~\cite{Plumari:2017ntm}, jet quenching mechanisms are not included in \pp collisions. The model predicts that hadronisation via coalescence is dominant at low \pt, while fragmentation dominates at high \pt. Both the magnitude and the \pt shape of the measured \LcD ratio are described well by this model.

\begin{figure}[t!b]
\centering
	\includegraphics[width=0.49\textwidth]{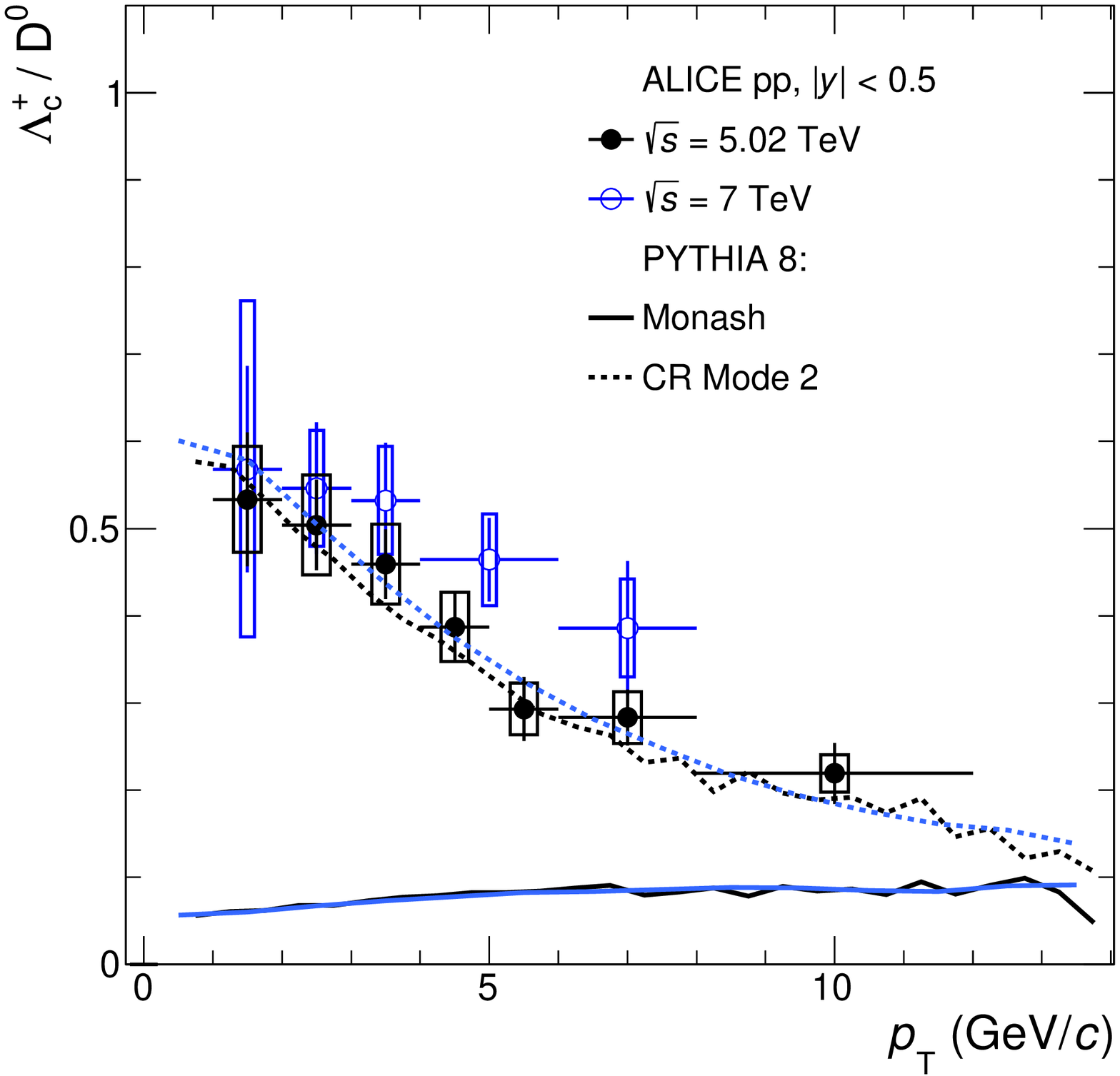}
	\includegraphics[width=0.49\textwidth]{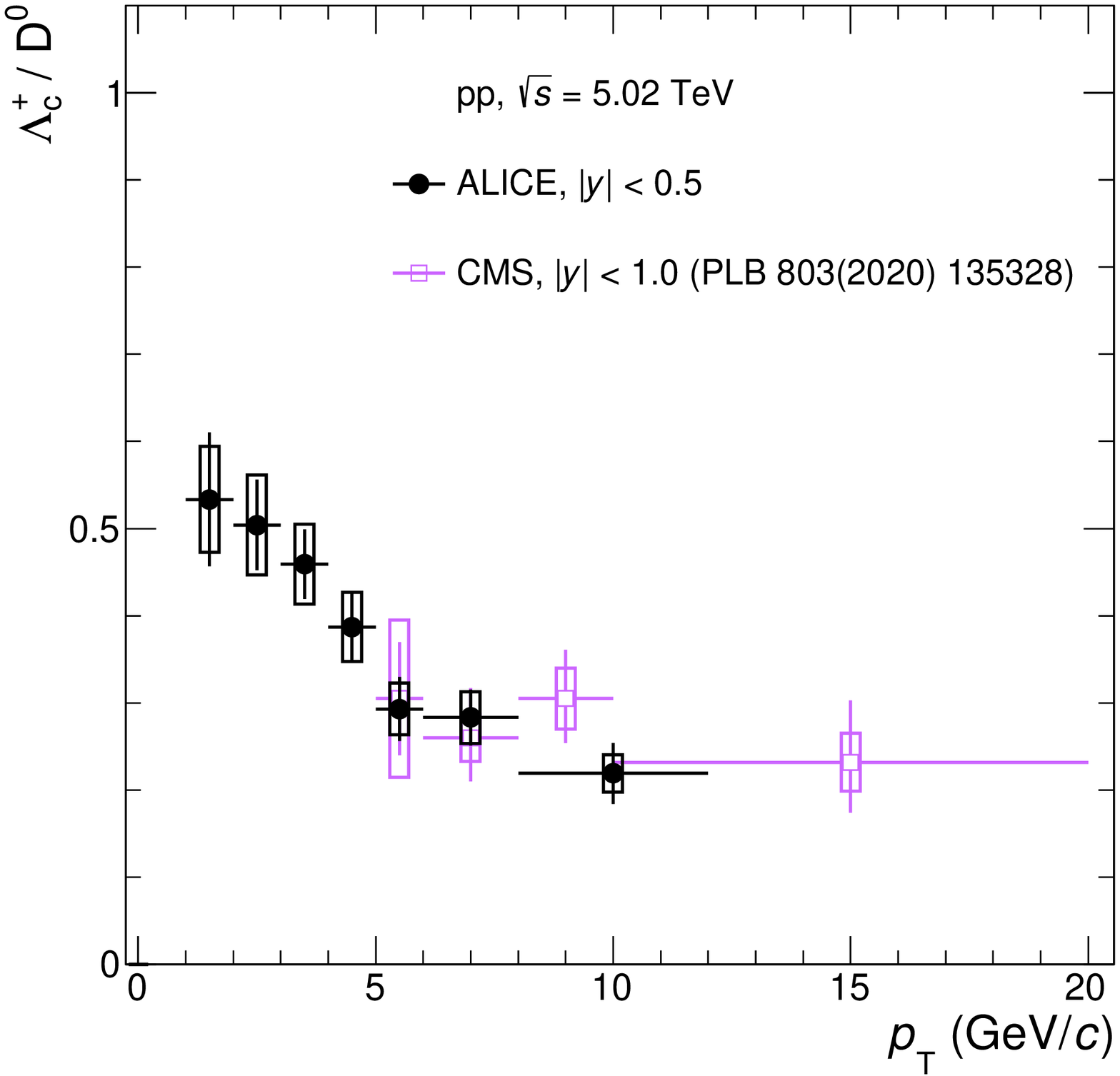}
\caption{Left: The \LcD ratio measured in pp collisions at \sqrtsfive, compared to the measurement at \sqrtsseven~\cite{Acharya:2017kfy}. PYTHIA 8 predictions are shown at both energies, for the Monash tune (solid lines) and with the Mode 2 CR tune (dotted lines). Right: the \LcD ratio at \sqrtsfive compared with the measurement by the CMS Collaboration at $|y|<1$~\cite{Sirunyan:2019fnc}.}
\label{fig:LcD-ppcompar}
\end{figure}

\Figref{fig:LcD-ppcompar} (left) shows the \LcD ratio in pp collisions at \sqrtsfive compared with the previous measurement at \sqrtsseven, and with predictions from PYTHIA 8 simulations. 
The \LcD ratio is found to be consistent between the two collision energies, within the experimental uncertainties; however, the wider \pt coverage and the improved statistical and systematic uncertainties on the new measurement reveal a clear decreasing trend in the \LcD ratio in pp collisions at \sqrtsfive, which was not clearly visible in the result at $\sqrts = 7~\tev$. The predictions of PYTHIA 8 with Monash tune do not show a $\sqrts$-dependence, while those with CR Mode 2 indicate a slight $\sqrts$-dependence, where the \LcD ratio is slightly larger at low \pt at $\sqrts = 7~\tev$ than at $\sqrts = 5.02~\tev$. The right panel shows the \LcD ratio in pp collisions, compared with the measurement by the CMS Collaboration in $5<\pt<20~\GeVc$ and $|y| < 1$~\cite{Sirunyan:2019fnc}. In the \pt region covered by both experiments, the results are found to be consistent with one another.

\begin{figure}[ht!]
	\centering
	\includegraphics[width=0.49\textwidth]{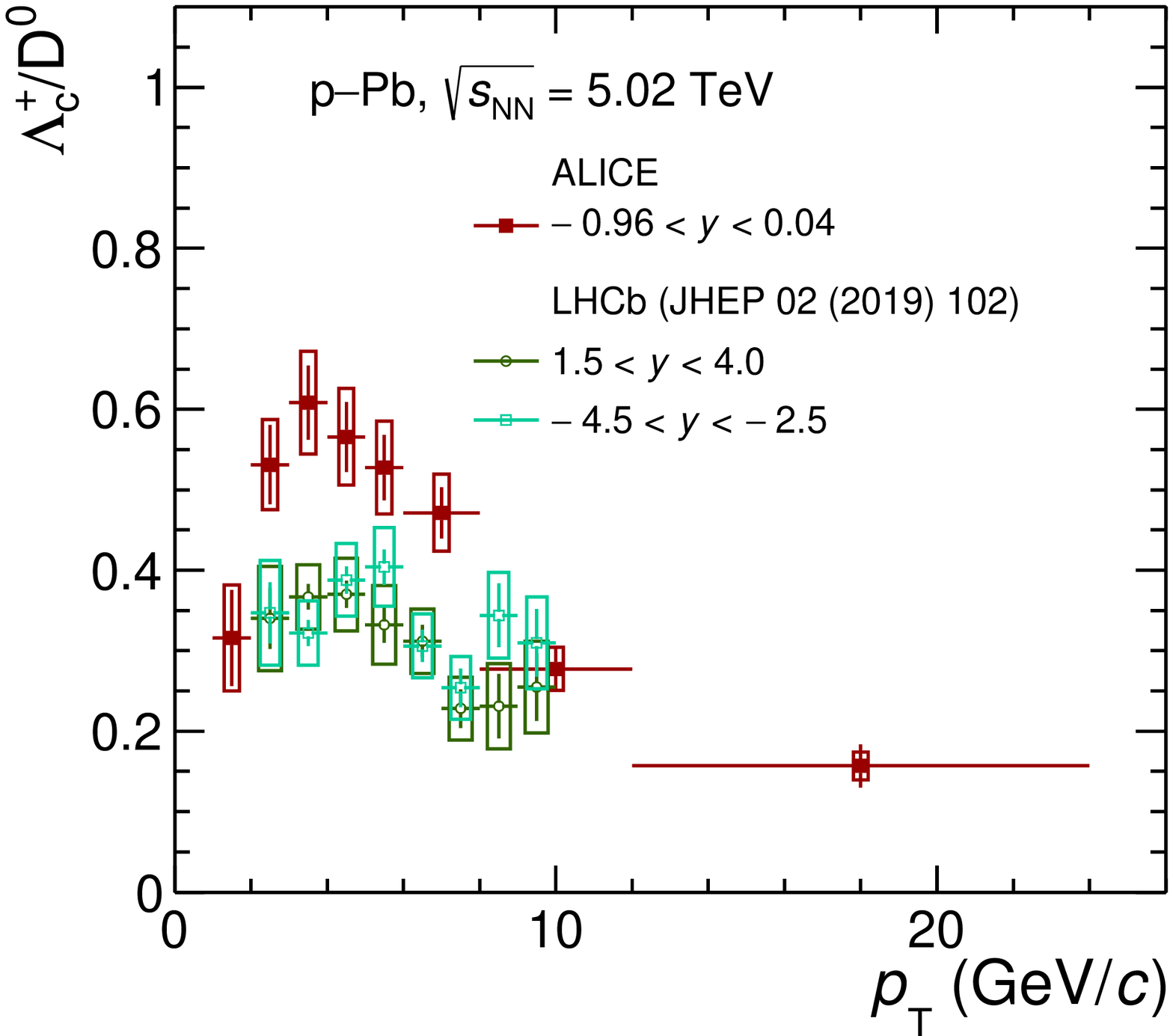}	
	\includegraphics[width=0.49\textwidth]{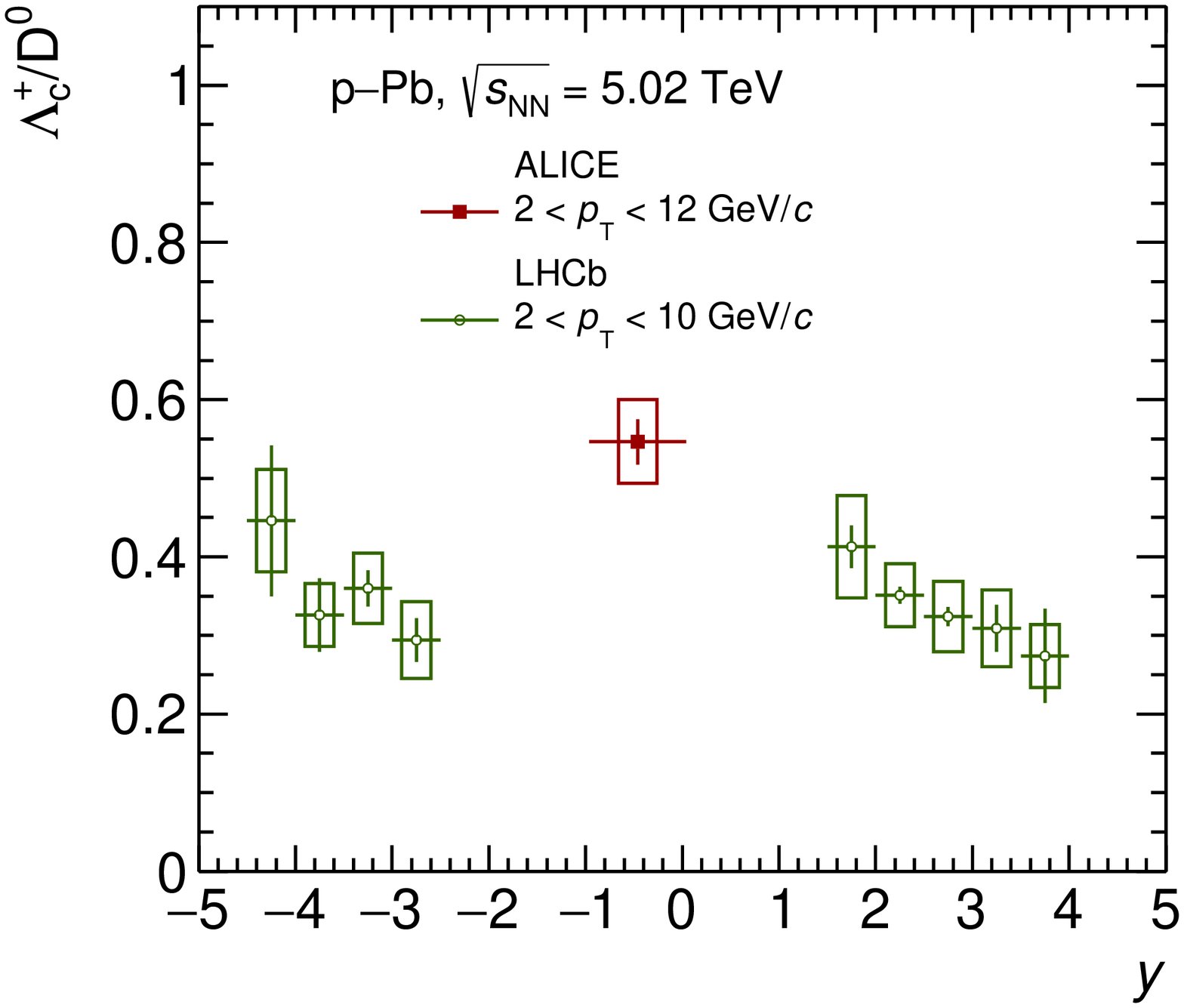}	
	\caption{The \LcD ratio measured in \pPb collisions at \sqrtsNNfive, compared with the measurement at forward and backward rapidity~\cite{Aaij:2018iyy} by the LHCb collaboration. The measurements are shown as a function of \pt (left) and as a function of $y$ (right).}
	\label{fig:LcD-LHCbComparison}
\end{figure}

In~\figref{fig:LcD-LHCbComparison}, the \LcD ratio in \pPb collisions at midrapidity ($-0.96<y<0.04$) is compared with the measurements by the LHCb Collaboration at forward ($1.5<y<4$) and backward (${-4.5<y<-2.5}$) rapidities~\cite{Aaij:2018iyy}. The left panel shows the comparison of the \LcD ratios in the different rapidity intervals as a function of \pt. For $\pt < 8$\,\gevc the ratio measured at midrapidity is higher than the ones measured at forward and backward rapidities, whereas at higher \pt the measurements are consistent within uncertainties. The right panel shows the \pt-integrated \LcD ratio as a function of rapidity. The \pt range of the integration of the ALICE data ($2 < \pt < 12$\,\gevc) is chosen to be similar to the reported LHCb integrated \pt range ($2 < \pt < 10$\,\gevc). The results suggest an enhancement of the ratio at midrapidity with respect to forward and backward rapidities. The difference between the \LcD ratio at mid and forward (backward) rapidities is less pronounced in \pPb collisions compared to the one observed in \pp collisions at $7~\tev$~\cite{Aaij:2013mga,Acharya:2017kfy}.

\begin{figure}[ht!]
	\centering
	\includegraphics[width=0.98\textwidth]{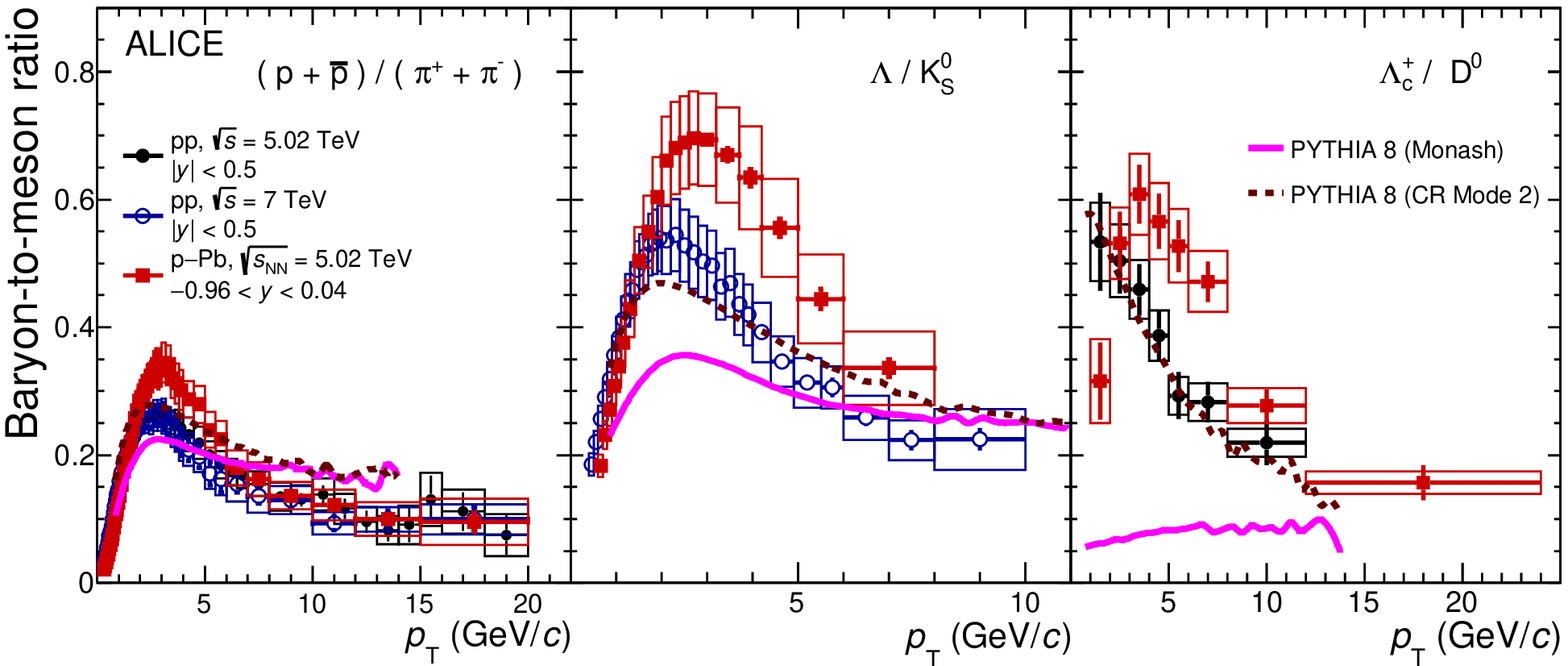}
	\caption{The baryon-to-meson ratios in the light-flavour and charm sector; p/$\pi$ in pp collisions at $\sqrts = 5.02~\tev$ and $7~\tev$ and \pPb collisions at $\sqrtsNN = 5.02~\tev$~\cite{Adam:2016dau} (left), $\Lambda/\Kzs$ in pp collisions at $\sqrts = 7~\tev$ and \pPb collisions at $\sqrtsNN = 5.02~\tev$~\cite{Abelev:2013xaa,Abelev:2013haa} (middle), and \LcD in \pp collisions at $\sqrts = 5.02~\tev$ and \pPb collisions at $\sqrtsNN = 5.02~\tev$ (right). The data are compared to predictions from PYTHIA 8~\cite{Skands:2014pea,Christiansen:2015yqa}. See text for model details.}
	\label{fig:LcD-LF}
\end{figure}

\Figref{fig:LcD-LF} shows the \LcD ratio in \pp and \pPb collisions, compared to the baryon-to-meson ratios in the light flavour sector, p/$\pi$~\cite{Adam:2016dau,Acharya:2019yoi} and $\Lambda/\Kzs$~\cite{Abelev:2013xaa,Abelev:2013haa}.
The p/$\pi$ ratio in pp collisions is shown at centre-of-mass energies of $7~\tev$ and $5.02~\tev$, and both results are fully consistent with each other. The $\Lambda/\Kzs$ ratio in \pp collisions is shown at $\sqrts = 7~\tev$.
Comparing the \LcD ratio to the light-flavour ratios, similar characteristics can be seen. All the baryon-to-meson ratios decrease with increasing \pt for $\pt > 3$ \gevc. In addition, the light-flavour hadron ratios show a distinct peak at intermediate \pt (around 3 \gevc), while the \LcD ratio shows a hint of a peak at $2 < \pt < 4$ \gevc in \pPb collisions, though a higher precision measurement would be needed to confirm this.  Also shown in \figref{fig:LcD-LF} are predictions from PYTHIA 8 with Monash and CR Mode 2 tunes. 
The PYTHIA 8 predictions for the light-flavour baryon-to-meson ratios are calculated at $\sqrts = 7~\tev$. 
It can be observed that the behaviours of the PYTHIA 8 predictions for light-flavour and charm baryon-to-meson ratios are similar. 
The measured $\Lambda / \Kzs$ ratio in pp collisions is underestimated by the Monash tune, while for the CR Mode 2 tune both the magnitude and trend of the ratio are closer to data, despite predicting a slightly flatter trend with \pt. The p$/\pi$ ratio is underestimated by PYTHIA 8 (Monash) at low \pt but overestimated at high \pt, while CR Mode 2 improves the agreement with data at low \pt but still overestimates the data at high \pt. 
Overall, the colour reconnection modes in \mbox{PYTHIA 8} generally provide a better description of the baryon-to-meson ratios in both the light-flavour and charm sector.

\section{Summary and conclusions}
\label{sec: Conclusions}
The measurements of the production of prompt \Lc baryons at midrapidity in \pp collisions at $\sqrts = 5.02~\tev$ and in \pPb collisions at $\sqrtsNN = 5.02~\tev$ with the ALICE detector at the LHC have been reported. The measurement in pp collisions, in particular, was performed at a different centre-of-mass energy with respect to the previous work in which \Lc-baryon production was measured in pp collisions at \sqrtsseven~\cite{Acharya:2017kfy}. The pp data sample at \sqrtsfive is the natural reference for measurements in \pPb and \PbPb collisions at the same centre-of-mass energy per nucleon pair. Moreover, with respect to~\cite{Acharya:2017kfy}, the uncertainties were significantly reduced, and the \pT range and the \pT granularity of the measurements were improved in both collision systems.
The analysis was performed using two different decay channels, \LcpKpi and \LctopKzS.
The results were reported for \pp collisions in the rapidity interval $|y| < 0.5$ and the transverse-momentum interval $1<\pT<12$~\gevc and for \pPb collisions in $-0.96 < y < 0.04$ and $1<\pT<24$~\gevc.
The \pT-differential production cross sections were obtained averaging the results from different hadronic decay channels. 

The \pt-differential cross section was measured to be larger than predictions given by pQCD calculations in both \pp and \pPb collisions.
The nuclear modification factor \RpPb~of \Lc baryons was found to be below unity in the interval $1 < \pT < 2~\gevc$ and to peak above unity around $5~\gevc$. It is consistent with the \RpPb~of D mesons in the \pt regions $1 < \pt< 4$ \gevc and $\pt>8$ \gevc and larger than the D-meson \RpPb in $4 < \pt < 8~\gevc$. The current precision of the measurement is not enough to draw conclusions on the role of different CNM effects and the possible presence of hot-medium effects.
As already observed in~\cite{Acharya:2017kfy}, the \LcD baryon-to-meson ratio in \pp collisions is larger than previous measurements obtained in $\ee$ and $\ep$ collision systems at lower centre-of-mass energies. The increase of precision in this paper allowed to observe, for the first time, a clear decreasing trend as a function of transverse momentum in the \LcD ratio. 
The \LcD ratio was compared to \pp event generators and models that implement different particle production and hadronisation mechanisms: qualitative agreement with the measurement is obtained with \mbox{PYTHIA 8} tunes including string formation beyond the leading-colour approximation; a prediction based on the statistical hadronisation model which includes unobserved charm baryon states that strongly decay to \Lc; and a prediction which assumes the formation of a QGP and implements hadronisation via coalescence and fragmentation. 
The \LcD ratio measured in \pp collisions is consistent with the results by CMS at midrapidity in the common \pt regions of both measurements. The ratio in \pPb collisions at midrapidity is higher than the one measured by LHCb at forward and backward rapidities in $2 < \pt < 8~\gevc$, while for $\pt>8~\gevc$ the measurements at central, forward and backward rapidities are consistent within uncertainties.
The measured \LcD ratio was also compared with baryon-to-meson ratios measured in the light-flavour sector. 
The measured $\Lambda / \Kzs$ ratio can also be described by PYTHIA 8 when including string formation beyond the leading-colour approximation, although this \mbox{PYTHIA 8} tune slightly overestimates the measured p$/\pi$ ratio.
The increased precision of this measurement with respect to the measurements
made with the Run 1 data is crucial for providing further insight into charm baryon production in \pp and \pPb collisions.
A more precise measurement is expected to be obtained during the LHC Run 3 and Run 4 after the upgrade of the ALICE apparatus~\cite{Abelevetal:2014cna}.  

\newenvironment{acknowledgement}{\relax}{\relax}
\begin{acknowledgement}
\section*{Acknowledgements}

The ALICE Collaboration would like to thank all its engineers and technicians for their invaluable contributions to the construction of the experiment and the CERN accelerator teams for the outstanding performance of the LHC complex.
The ALICE Collaboration gratefully acknowledges the resources and support provided by all Grid centres and the Worldwide LHC Computing Grid (WLCG) collaboration.
The ALICE Collaboration acknowledges the following funding agencies for their support in building and running the ALICE detector:
A. I. Alikhanyan National Science Laboratory (Yerevan Physics Institute) Foundation (ANSL), State Committee of Science and World Federation of Scientists (WFS), Armenia;
Austrian Academy of Sciences, Austrian Science Fund (FWF): [M 2467-N36] and Nationalstiftung f\"{u}r Forschung, Technologie und Entwicklung, Austria;
Ministry of Communications and High Technologies, National Nuclear Research Center, Azerbaijan;
Conselho Nacional de Desenvolvimento Cient\'{\i}fico e Tecnol\'{o}gico (CNPq), Financiadora de Estudos e Projetos (Finep), Funda\c{c}\~{a}o de Amparo \`{a} Pesquisa do Estado de S\~{a}o Paulo (FAPESP) and Universidade Federal do Rio Grande do Sul (UFRGS), Brazil;
Ministry of Education of China (MOEC) , Ministry of Science \& Technology of China (MSTC) and National Natural Science Foundation of China (NSFC), China;
Ministry of Science and Education and Croatian Science Foundation, Croatia;
Centro de Aplicaciones Tecnol\'{o}gicas y Desarrollo Nuclear (CEADEN), Cubaenerg\'{\i}a, Cuba;
Ministry of Education, Youth and Sports of the Czech Republic, Czech Republic;
The Danish Council for Independent Research | Natural Sciences, the VILLUM FONDEN and Danish National Research Foundation (DNRF), Denmark;
Helsinki Institute of Physics (HIP), Finland;
Commissariat \`{a} l'Energie Atomique (CEA) and Institut National de Physique Nucl\'{e}aire et de Physique des Particules (IN2P3) and Centre National de la Recherche Scientifique (CNRS), France;
Bundesministerium f\"{u}r Bildung und Forschung (BMBF) and GSI Helmholtzzentrum f\"{u}r Schwerionenforschung GmbH, Germany;
General Secretariat for Research and Technology, Ministry of Education, Research and Religions, Greece;
National Research, Development and Innovation Office, Hungary;
Department of Atomic Energy Government of India (DAE), Department of Science and Technology, Government of India (DST), University Grants Commission, Government of India (UGC) and Council of Scientific and Industrial Research (CSIR), India;
Indonesian Institute of Science, Indonesia;
Istituto Nazionale di Fisica Nucleare (INFN), Italy;
Institute for Innovative Science and Technology , Nagasaki Institute of Applied Science (IIST), Japanese Ministry of Education, Culture, Sports, Science and Technology (MEXT) and Japan Society for the Promotion of Science (JSPS) KAKENHI, Japan;
Consejo Nacional de Ciencia (CONACYT) y Tecnolog\'{i}a, through Fondo de Cooperaci\'{o}n Internacional en Ciencia y Tecnolog\'{i}a (FONCICYT) and Direcci\'{o}n General de Asuntos del Personal Academico (DGAPA), Mexico;
Nederlandse Organisatie voor Wetenschappelijk Onderzoek (NWO), Netherlands;
The Research Council of Norway, Norway;
Commission on Science and Technology for Sustainable Development in the South (COMSATS), Pakistan;
Pontificia Universidad Cat\'{o}lica del Per\'{u}, Peru;
Ministry of Science and Higher Education, National Science Centre and WUT ID-UB, Poland;
Korea Institute of Science and Technology Information and National Research Foundation of Korea (NRF), Republic of Korea;
Ministry of Education and Scientific Research, Institute of Atomic Physics and Ministry of Research and Innovation and Institute of Atomic Physics, Romania;
Joint Institute for Nuclear Research (JINR), Ministry of Education and Science of the Russian Federation, National Research Centre Kurchatov Institute, Russian Science Foundation and Russian Foundation for Basic Research, Russia;
Ministry of Education, Science, Research and Sport of the Slovak Republic, Slovakia;
National Research Foundation of South Africa, South Africa;
Swedish Research Council (VR) and Knut \& Alice Wallenberg Foundation (KAW), Sweden;
European Organization for Nuclear Research, Switzerland;
Suranaree University of Technology (SUT), National Science and Technology Development Agency (NSDTA) and Office of the Higher Education Commission under NRU project of Thailand, Thailand;
Turkish Atomic Energy Agency (TAEK), Turkey;
National Academy of  Sciences of Ukraine, Ukraine;
Science and Technology Facilities Council (STFC), United Kingdom;
National Science Foundation of the United States of America (NSF) and United States Department of Energy, Office of Nuclear Physics (DOE NP), United States of America.
\end{acknowledgement}

\bibliographystyle{utphys}   
\bibliography{include/bibfile}

\newpage
\appendix

%
%

\section{The ALICE Collaboration}
\label{app:collab}
\begingroup
\small
\begin{flushleft}


S.~Acharya$^{\rm 142}$, 
D.~Adamov\'{a}$^{\rm 97}$, 
A.~Adler$^{\rm 75}$, 
J.~Adolfsson$^{\rm 82}$, 
G.~Aglieri Rinella$^{\rm 35}$, 
M.~Agnello$^{\rm 31}$, 
N.~Agrawal$^{\rm 55}$, 
Z.~Ahammed$^{\rm 142}$, 
S.~Ahmad$^{\rm 16}$, 
S.U.~Ahn$^{\rm 77}$
% ...

\section*{Affiliation notes}

$^{\rm I}$ Deceased\\
$^{\rm II}$ Also at: Italian National Agency for New Technologies, Energy and Sustainable Economic Development (ENEA), Bologna, Italy\\
$^{\rm III}$ Also at: Dipartimento DET del Politecnico di Torino, Turin, Italy\\
$^{\rm IV}$ Also at: M.V. Lomonosov Moscow State University, D.V. Skobeltsyn Institute of Nuclear, Physics, Moscow, Russia\\
$^{\rm V}$ Also at: Institute of Theoretical Physics, University of Wroclaw, Poland\\

\section*{Collaboration Institutes}

$^{1}$ A.I. Alikhanyan National Science Laboratory (Yerevan Physics Institute) Foundation, Yerevan, Armenia\\
$^{2}$ AGH University of Science and Technology, Cracow, Poland\\
$^{3}$ Bogolyubov Institute for Theoretical Physics, National Academy of Sciences of Ukraine, Kiev, Ukraine\\
$^{4}$ Bose Institute, Department of Physics  and Centre for Astroparticle Physics and Space Science (CAPSS), Kolkata, India\\
$^{5}$ Budker Institute for Nuclear Physics, Novosibirsk, Russia\\
$^{6}$ California Polytechnic State University, San Luis Obispo, California, United States\\
$^{7}$ Central China Normal University, Wuhan, China\\
$^{8}$ Centro de Aplicaciones Tecnol\'{o}gicas y Desarrollo Nuclear (CEADEN), Havana, Cuba\\
$^{9}$ Centro de Investigaci\'{o}n y de Estudios Avanzados (CINVESTAV), Mexico City and M\'{e}rida, Mexico\\
$^{10}$ Chicago State University, Chicago, Illinois, United States\\
$^{11}$ China Institute of Atomic Energy, Beijing, China\\
$^{12}$ Chungbuk National University, Cheongju, Republic of Korea\\
$^{13}$ Comenius University Bratislava, Faculty of Mathematics, Physics and Informatics, Bratislava, Slovakia\\
$^{14}$ COMSATS University Islamabad, Islamabad, Pakistan\\
$^{15}$ Creighton University, Omaha, Nebraska, United States\\
$^{16}$ Department of Physics, Aligarh Muslim University, Aligarh, India\\
$^{17}$ Department of Physics, Pusan National University, Pusan, Republic of Korea\\
$^{18}$ Department of Physics, Sejong University, Seoul, Republic of Korea\\
$^{19}$ Department of Physics, University of California, Berkeley, California, United States\\
$^{20}$ Department of Physics, University of Oslo, Oslo, Norway\\
$^{21}$ Department of Physics and Technology, University of Bergen, Bergen, Norway\\
$^{22}$ Dipartimento di Fisica dell'Universit\`{a} 'La Sapienza' and Sezione INFN, Rome, Italy\\
$^{23}$ Dipartimento di Fisica dell'Universit\`{a} and Sezione INFN, Cagliari, Italy\\
$^{24}$ Dipartimento di Fisica dell'Universit\`{a} and Sezione INFN, Trieste, Italy\\
$^{25}$ Dipartimento di Fisica dell'Universit\`{a} and Sezione INFN, Turin, Italy\\
$^{26}$ Dipartimento di Fisica e Astronomia dell'Universit\`{a} and Sezione INFN, Bologna, Italy\\
$^{27}$ Dipartimento di Fisica e Astronomia dell'Universit\`{a} and Sezione INFN, Catania, Italy\\
$^{28}$ Dipartimento di Fisica e Astronomia dell'Universit\`{a} and Sezione INFN, Padova, Italy\\
$^{29}$ Dipartimento di Fisica e Nucleare e Teorica, Universit\`{a} di Pavia  and Sezione INFN, Pavia, Italy\\
$^{30}$ Dipartimento di Fisica `E.R.~Caianiello' dell'Universit\`{a} and Gruppo Collegato INFN, Salerno, Italy\\
$^{31}$ Dipartimento DISAT del Politecnico and Sezione INFN, Turin, Italy\\
$^{32}$ Dipartimento di Scienze e Innovazione Tecnologica dell'Universit\`{a} del Piemonte Orientale and INFN Sezione di Torino, Alessandria, Italy\\
$^{33}$ Dipartimento di Scienze MIFT, Universit\`{a} di Messina, Messina, Italy\\
$^{34}$ Dipartimento Interateneo di Fisica `M.~Merlin' and Sezione INFN, Bari, Italy\\
$^{35}$ European Organization for Nuclear Research (CERN), Geneva, Switzerland\\
$^{36}$ Faculty of Electrical Engineering, Mechanical Engineering and Naval Architecture, University of Split, Split, Croatia\\
$^{37}$ Faculty of Engineering and Science, Western Norway University of Applied Sciences, Bergen, Norway\\
$^{38}$ Faculty of Nuclear Sciences and Physical Engineering, Czech Technical University in Prague, Prague, Czech Republic\\
$^{39}$ Faculty of Science, P.J.~\v{S}af\'{a}rik University, Ko\v{s}ice, Slovakia\\
$^{40}$ Frankfurt Institute for Advanced Studies, Johann Wolfgang Goethe-Universit\"{a}t Frankfurt, Frankfurt, Germany\\
$^{41}$ Fudan University, Shanghai, China\\
$^{42}$ Gangneung-Wonju National University, Gangneung, Republic of Korea\\
$^{43}$ Gauhati University, Department of Physics, Guwahati, India\\
$^{44}$ Helmholtz-Institut f\"{u}r Strahlen- und Kernphysik, Rheinische Friedrich-Wilhelms-Universit\"{a}t Bonn, Bonn, Germany\\
$^{45}$ Helsinki Institute of Physics (HIP), Helsinki, Finland\\
$^{46}$ High Energy Physics Group,  Universidad Aut\'{o}noma de Puebla, Puebla, Mexico\\
$^{47}$ Hiroshima University, Hiroshima, Japan\\
$^{48}$ Hochschule Worms, Zentrum  f\"{u}r Technologietransfer und Telekommunikation (ZTT), Worms, Germany\\
$^{49}$ Horia Hulubei National Institute of Physics and Nuclear Engineering, Bucharest, Romania\\
$^{50}$ Indian Institute of Technology Bombay (IIT), Mumbai, India\\
$^{51}$ Indian Institute of Technology Indore, Indore, India\\
$^{52}$ Indonesian Institute of Sciences, Jakarta, Indonesia\\
$^{53}$ INFN, Laboratori Nazionali di Frascati, Frascati, Italy\\
$^{54}$ INFN, Sezione di Bari, Bari, Italy\\
$^{55}$ INFN, Sezione di Bologna, Bologna, Italy\\
$^{56}$ INFN, Sezione di Cagliari, Cagliari, Italy\\
$^{57}$ INFN, Sezione di Catania, Catania, Italy\\
$^{58}$ INFN, Sezione di Padova, Padova, Italy\\
$^{59}$ INFN, Sezione di Roma, Rome, Italy\\
$^{60}$ INFN, Sezione di Torino, Turin, Italy\\
$^{61}$ INFN, Sezione di Trieste, Trieste, Italy\\
$^{62}$ Inha University, Incheon, Republic of Korea\\
$^{63}$ Institute for Gravitational and Subatomic Physics (GRASP), Utrecht University/Nikhef, Utrecht, Netherlands\\
$^{64}$ Institute for Nuclear Research, Academy of Sciences, Moscow, Russia\\
$^{65}$ Institute of Experimental Physics, Slovak Academy of Sciences, Ko\v{s}ice, Slovakia\\
$^{66}$ Institute of Physics, Homi Bhabha National Institute, Bhubaneswar, India\\
$^{67}$ Institute of Physics of the Czech Academy of Sciences, Prague, Czech Republic\\
$^{68}$ Institute of Space Science (ISS), Bucharest, Romania\\
$^{69}$ Institut f\"{u}r Kernphysik, Johann Wolfgang Goethe-Universit\"{a}t Frankfurt, Frankfurt, Germany\\
$^{70}$ Instituto de Ciencias Nucleares, Universidad Nacional Aut\'{o}noma de M\'{e}xico, Mexico City, Mexico\\
$^{71}$ Instituto de F\'{i}sica, Universidade Federal do Rio Grande do Sul (UFRGS), Porto Alegre, Brazil\\
$^{72}$ Instituto de F\'{\i}sica, Universidad Nacional Aut\'{o}noma de M\'{e}xico, Mexico City, Mexico\\
$^{73}$ iThemba LABS, National Research Foundation, Somerset West, South Africa\\
$^{74}$ Jeonbuk National University, Jeonju, Republic of Korea\\
$^{75}$ Johann-Wolfgang-Goethe Universit\"{a}t Frankfurt Institut f\"{u}r Informatik, Fachbereich Informatik und Mathematik, Frankfurt, Germany\\
$^{76}$ Joint Institute for Nuclear Research (JINR), Dubna, Russia\\
$^{77}$ Korea Institute of Science and Technology Information, Daejeon, Republic of Korea\\
$^{78}$ KTO Karatay University, Konya, Turkey\\
$^{79}$ Laboratoire de Physique des 2 Infinis, Ir\`{e}ne Joliot-Curie, Orsay, France\\
$^{80}$ Laboratoire de Physique Subatomique et de Cosmologie, Universit\'{e} Grenoble-Alpes, CNRS-IN2P3, Grenoble, France\\
$^{81}$ Lawrence Berkeley National Laboratory, Berkeley, California, United States\\
$^{82}$ Lund University Department of Physics, Division of Particle Physics, Lund, Sweden\\
$^{83}$ Moscow Institute for Physics and Technology, Moscow, Russia\\
$^{84}$ Nagasaki Institute of Applied Science, Nagasaki, Japan\\
$^{85}$ Nara Women{'}s University (NWU), Nara, Japan\\
$^{86}$ National and Kapodistrian University of Athens, School of Science, Department of Physics , Athens, Greece\\
$^{87}$ National Centre for Nuclear Research, Warsaw, Poland\\
$^{88}$ National Institute of Science Education and Research, Homi Bhabha National Institute, Jatni, India\\
$^{89}$ National Nuclear Research Center, Baku, Azerbaijan\\
$^{90}$ National Research Centre Kurchatov Institute, Moscow, Russia\\
$^{91}$ Niels Bohr Institute, University of Copenhagen, Copenhagen, Denmark\\
$^{92}$ Nikhef, National institute for subatomic physics, Amsterdam, Netherlands\\
$^{93}$ NRC Kurchatov Institute IHEP, Protvino, Russia\\
$^{94}$ NRC \guillemotleft Kurchatov\guillemotright  Institute - ITEP, Moscow, Russia\\
$^{95}$ NRNU Moscow Engineering Physics Institute, Moscow, Russia\\
$^{96}$ Nuclear Physics Group, STFC Daresbury Laboratory, Daresbury, United Kingdom\\
$^{97}$ Nuclear Physics Institute of the Czech Academy of Sciences, \v{R}e\v{z} u Prahy, Czech Republic\\
$^{98}$ Oak Ridge National Laboratory, Oak Ridge, Tennessee, United States\\
$^{99}$ Ohio State University, Columbus, Ohio, United States\\
$^{100}$ Petersburg Nuclear Physics Institute, Gatchina, Russia\\
$^{101}$ Physics department, Faculty of science, University of Zagreb, Zagreb, Croatia\\
$^{102}$ Physics Department, Panjab University, Chandigarh, India\\
$^{103}$ Physics Department, University of Jammu, Jammu, India\\
$^{104}$ Physics Department, University of Rajasthan, Jaipur, India\\
$^{105}$ Physikalisches Institut, Eberhard-Karls-Universit\"{a}t T\"{u}bingen, T\"{u}bingen, Germany\\
$^{106}$ Physikalisches Institut, Ruprecht-Karls-Universit\"{a}t Heidelberg, Heidelberg, Germany\\
$^{107}$ Physik Department, Technische Universit\"{a}t M\"{u}nchen, Munich, Germany\\
$^{108}$ Politecnico di Bari and Sezione INFN, Bari, Italy\\
$^{109}$ Research Division and ExtreMe Matter Institute EMMI, GSI Helmholtzzentrum f\"ur Schwerionenforschung GmbH, Darmstadt, Germany\\
$^{110}$ Rudjer Bo\v{s}kovi\'{c} Institute, Zagreb, Croatia\\
$^{111}$ Russian Federal Nuclear Center (VNIIEF), Sarov, Russia\\
$^{112}$ Saha Institute of Nuclear Physics, Homi Bhabha National Institute, Kolkata, India\\
$^{113}$ School of Physics and Astronomy, University of Birmingham, Birmingham, United Kingdom\\
$^{114}$ Secci\'{o}n F\'{\i}sica, Departamento de Ciencias, Pontificia Universidad Cat\'{o}lica del Per\'{u}, Lima, Peru\\
$^{115}$ St. Petersburg State University, St. Petersburg, Russia\\
$^{116}$ Stefan Meyer Institut f\"{u}r Subatomare Physik (SMI), Vienna, Austria\\
$^{117}$ SUBATECH, IMT Atlantique, Universit\'{e} de Nantes, CNRS-IN2P3, Nantes, France\\
$^{118}$ Suranaree University of Technology, Nakhon Ratchasima, Thailand\\
$^{119}$ Technical University of Ko\v{s}ice, Ko\v{s}ice, Slovakia\\
$^{120}$ The Henryk Niewodniczanski Institute of Nuclear Physics, Polish Academy of Sciences, Cracow, Poland\\
$^{121}$ The University of Texas at Austin, Austin, Texas, United States\\
$^{122}$ Universidad Aut\'{o}noma de Sinaloa, Culiac\'{a}n, Mexico\\
$^{123}$ Universidade de S\~{a}o Paulo (USP), S\~{a}o Paulo, Brazil\\
$^{124}$ Universidade Estadual de Campinas (UNICAMP), Campinas, Brazil\\
$^{125}$ Universidade Federal do ABC, Santo Andre, Brazil\\
$^{126}$ University of Cape Town, Cape Town, South Africa\\
$^{127}$ University of Houston, Houston, Texas, United States\\
$^{128}$ University of Jyv\"{a}skyl\"{a}, Jyv\"{a}skyl\"{a}, Finland\\
$^{129}$ University of Liverpool, Liverpool, United Kingdom\\
$^{130}$ University of Science and Technology of China, Hefei, China\\
$^{131}$ University of South-Eastern Norway, Tonsberg, Norway\\
$^{132}$ University of Tennessee, Knoxville, Tennessee, United States\\
$^{133}$ University of the Witwatersrand, Johannesburg, South Africa\\
$^{134}$ University of Tokyo, Tokyo, Japan\\
$^{135}$ University of Tsukuba, Tsukuba, Japan\\
$^{136}$ Universit\'{e} Clermont Auvergne, CNRS/IN2P3, LPC, Clermont-Ferrand, France\\
$^{137}$ Universit\'{e} de Lyon, CNRS/IN2P3, Institut de Physique des 2 Infinis de Lyon , Lyon, France\\
$^{138}$ Universit\'{e} de Strasbourg, CNRS, IPHC UMR 7178, F-67000 Strasbourg, France, Strasbourg, France\\
$^{139}$ Universit\'{e} Paris-Saclay Centre d'Etudes de Saclay (CEA), IRFU, D\'{e}partment de Physique Nucl\'{e}aire (DPhN), Saclay, France\\
$^{140}$ Universit\`{a} degli Studi di Foggia, Foggia, Italy\\
$^{141}$ Universit\`{a} di Brescia and Sezione INFN, Brescia, Italy\\
$^{142}$ Variable Energy Cyclotron Centre, Homi Bhabha National Institute, Kolkata, India\\
$^{143}$ Warsaw University of Technology, Warsaw, Poland\\
$^{144}$ Wayne State University, Detroit, Michigan, United States\\
$^{145}$ Westf\"{a}lische Wilhelms-Universit\"{a}t M\"{u}nster, Institut f\"{u}r Kernphysik, M\"{u}nster, Germany\\
$^{146}$ Wigner Research Centre for Physics, Budapest, Hungary\\
$^{147}$ Yale University, New Haven, Connecticut, United States\\
$^{148}$ Yonsei University, Seoul, Republic of Korea\\

\endgroup  
\end{document}